\documentclass[a4paper]{JHEP3} 
\usepackage{microtype}
\usepackage{lmodern}
\usepackage[T1]{fontenc}
\usepackage{color} 
\let\normalcolor\relax 

\usepackage{graphicx}
\usepackage{dcolumn}
\usepackage{bm}
\usepackage{amssymb}
\usepackage{mathrsfs}
\usepackage{amsmath}
\usepackage{epsfig}

\usepackage{amsmath}
\usepackage{rotating}
\usepackage{timestamp} 

\usepackage{epsf}
\def\be{\begin{equation}}
\def\ee{\end{equation}}
\def\bea{\begin{eqnarray}}
\def\eea{\end{eqnarray}}
\def\gsim{\ \rlap{\raise 2pt\hbox{$>$}}{\lower 2pt \hbox{$\sim$}}\ }
\def\lsim{\ \rlap{\raise 2pt\hbox{$<$}}{\lower 2pt \hbox{$\sim$}}\ }
\def\dslash{\kern-4pt \not{\hbox{\kern-2pt $\partial$}}}
\def\pslash{\not{\hbox{\kern-2pt p}}}

\title{Two Zero Mass Matrices and Sterile Neutrinos}
\author{Monojit Ghosh $^a$, Srubabati Goswami $^a$, Shivani Gupta $^b$\\
$^a$~Physical Research Laboratory, Navrangpura,
Ahmedabad 380 009, India \\
$^b$~Department of Physics and IPAP, Yonsei University, Seoul 120-479, Korea \\
Email: \email{monojit@prl.res.in, sruba@prl.res.in, shivani@cskim.yonsei.ac.kr}
}
\abstract
{Recent experimental data
is indicative of the existence of sterile neutrinos.
The minimal scheme that can account for the data and is
consistent with cosmological observations is
the 3+1 picture which consists of three predominantly
active and one predominantly sterile neutrino
with the fourth neutrino being heavier than the other three.
Within this scheme
there are two possibilities depending on whether the three light states
obey normal or inverted hierarchy.
In this paper we consider the two zero textures of the low energy
neutrino mass matrix
in presence of one additional sterile neutrino. We find that among 45 possible two zero
textures for this case,
15 are consistent with all current observations. Remarkably, these
correspond to the two-zero textures of a three active neutrino
mass matrix. We discuss the mass spectrum and the parameter
correlations that we find in the various textures. We also present
the effective mass
governing neutrinoless double beta decay as a function of the
lowest mass.
}

\begin{document}
\maketitle


\section{Introduction}

Neutrino oscillation in standard three flavour picture is
now well established from solar, atmospheric, reactor
and accelerator neutrinos. The mass squared differences governing
these oscillations are $\sim 10^{-4}$ eV$^2$ and $10^{-3}$ eV$^2$.
However, the reported observations of
$\bar{\nu}_\mu - \bar{\nu}_e$ oscillations
in LSND experiment \cite{lsnd} and recent confirmation of this
by the MiniBooNE experiment \cite{miniboone} with oscillation frequency
governed by a mass squared difference $\sim$ eV$^2$
cannot be accounted for in the above framework.
These  results motivate the introduction of atleast one extra
neutrino of mass $\sim$ eV to account for the three independent mass scales
governing solar, atmospheric and LSND oscillations.
LEP data on measurement of Z-line shape dictates that
there can be only three neutrinos with standard weak interactions and so
the fourth light neutrino, if it exists must be a Standard Model
singlet or sterile.
Recently this hypothesis  garnered additional
support from (i) disappearance of electron antineutrinos
in reactor experiments with recalculated
fluxes \cite{reactor} and (ii) deficit of  electron neutrinos
measured in the solar neutrino detectors GALLEX and SAGE
using radioactive sources \cite{ga}.
The  recent ICARUS results \cite{icarus} however, did not find any
evidence for the LSND oscillations.
But this does not completely rule out the LSND parameter space
and small active-sterile mixing still remains allowed
\cite{giunti_NPB}. Thus, the situation with sterile neutrinos
remains quite intriguing and many future experiments are proposed/planned to
test these results and reach a definitive conclusion
\cite{sterile-future}.

Addition of one extra sterile neutrino to the standard three
generation picture gives rise to two possible mass patterns --
the 2+2 and 3+1 scenarios \cite{earlier-papers}. Of these,
the 2+2 schemes are strongly disfavored
by the solar and atmospheric neutrino oscillation data \cite{valle}.
The 3+1 picture also suffers from some tension
between observation of oscillations in antineutrino channel by
LSND and MiniBooNE and non-observation of oscillations in the
neutrino channels as well as in disappearance measurements.
However, it was shown recently in \cite{giunti_3+1} that a reasonable
goodness-of-fit can still be obtained.
Although introduction of more than one sterile neutrinos
may provide a better fit to the
neutrino oscillation data \cite{3+2_fits}, the 3+1 scheme
is considered to be minimal and to be more consistent with the
cosmological data \cite{serpico}. Very recently combined analysis of
cosmological and short baseline (SBL) data in the context of additional sterile
neutrinos have been performed in \cite{giunti12,monoj12}.
The analysis in \cite{giunti12} found a preference of
the 3+1 scenario over 3+2
while the analysis in \cite{monoj12} shows that the status of the
3+2 scenario depends on the cosmological data set
used and the fitting procedure and no conclusive statement can be made
regarding whether it is favoured or disallowed.
In fact the current cosmological observations
of an weakly interacting relativistic "dark radiation"
may actually prefer
an additional sterile neutrino \cite{cosmo4}.
If this radiation is attributed  to extra neutrino species
then the data gives the bound on the number of neutrinos as
$N_{eff} = 4.08 \pm 0.8$ at 95\% C.L. \cite{cosmo4}.

In this paper we consider the structure of the low energy
neutrino mass matrices
in presence of one extra sterile neutrino.
The low energy mass matrix in the flavour basis is now  complex
symmetric with
10 independent entries and can be expressed in general as,

\begin{equation}
M_\nu
=\left(
\begin{array}{cccc}
m_{ee} & m_{e\mu} & m_{e\tau} & m_{es} \\
m_{e \mu}& m_{ \mu \mu} & m_{\mu \tau}& m_{\mu s} \\
m_{e \tau }& m_{\mu \tau} & m_{\tau \tau}& m_{\tau s}\\
m_{es} & m_{ \mu s} & m_{\tau s} & m_{ss}
\end{array}
\right)
\label{mf}
\end{equation}

For three flavours the
last row and the column would be absent and the mass matrix would
contain 6 elements. In the context of
three generations, a very remarkable result was obtained in
\cite{frampton} that there can be at the most two zeros in the
low energy neutrino
mass matrix in the flavour basis. 
For the three neutrino mass matrix there can be 15  possible 
two zero texture structures.    
These are the same as those shown in Table \ref{Table:2zero}
for the four neutrino case after omitting
the fourth row and column.
Among these, only 7 textures corresponding to the A, B and C class
were found to be
compatible with the experimental data on neutrino oscillation \cite{frampton}.
Normal Hierarchy (NH)  was found to be allowed in all the textures
whereas Inverted Hierarchy (IH) and Quasi-Degenerate (QD) solutions 
were allowed 
for the B and C classes. 
Various aspects of the two zero textures in the low energy
neutrino mass matrix  have been examined in \cite{shivani,Two-Zero-papers}.
Recently, this analysis has been redone in
\cite{Kumar, xingnew,Meloni,Morisi,Grimus}
to take into account the recent results
including the measured values of mixing angle $\theta_{13}$
by the reactor experiments \cite{dc}.
The analysis including the latest data and allowing the
parameters to vary randomly in their
3$\sigma$ range shows that
all 7 textures of the original analysis in \cite{frampton}
remain allowed \cite{Grimus}.  However,
with the 1$\sigma$ range of parameters
the scenarios become more constrained.
With the oscillation parameters taken from
\cite{fogli} only A class with NH remain allowed while for
oscillation parameters
in \cite{forero}  the textures belonging to
$B_2$ and $B_4$ classes for IH and C for NH  get excluded \cite{Grimus}.
This demonstrates  that with precise determination of oscillation
parameters the allowed scenarios would become more constrained.
However, the situation may change altogether in presence of sterile
neutrinos.

In this paper, we examine how many two zero
textures are allowed by the current oscillation data  in the low energy neutrino mass matrix
when an extra sterile neutrino is present.
We assume the known oscillation parameters are normally distributed 
with the peak at the best-fit value and 1$\sigma$ error as the
width.
First we check the  status of the two zero texture solutions
in the context of three generation mass matrices by this procedure.
Then we check how much these conclusions change in the 3+1 scenario with one
additional sterile neutrino.
We also investigate if any new interesting correlations can be found
specially for the sterile mixing angles.
Finally, we discuss the changes expected in our result if the 
mass and mixing parameters are varied randomly in their 3$\sigma$ range instead 
of varying  them in a Gaussian distribution peaked at the best-fit value. 

Texture zero implies some of the elements are much smaller
than the other elements of the mass matrix. Analysis of texture zeros
puts restriction
on the nature of the mass spectrum and can give rise to correlations
between the mixing angles, masses and CP phases which may be confirmed
or falsified by experimental observations.
This can often help in understanding the underlying flavour symmetry
\cite{anjan-grimus}. 
In case neutrino mass is generated by seesaw mechanism
the texture zeros in the low energy mass matrix
can be useful in
identifying the possible high scale Yukawa matrices
\cite{highscaletexture,Barbieri,Guo}.

The plan of the paper goes as follows. In the next section we discuss
our formalism. Section III discusses the results.
We end in section IV with summary and conclusions.
\begin{table}
\begin{small}
\begin{tiny}
\begin{center}
\begin{small}
\begin{tabular}{|c|c|c|c|}
\hline $ A_1$& $A_2$ &  &   \\
\hline $\left(
\begin{array}{cccc}
0 & 0 & \times &\times \\  0 & \times & \times & \times \\ \times & \times & \times & \times \\\times & \times & \times & \times
\end{array}
\right)$ & $\left(
\begin{array}{cccc}
0 & \times & 0 &\times \\  \times & \times & \times & \times \\ 0 & \times & \times & \times \\\times & \times & \times & \times
\end{array}
\right)$  & & \\
\hline $ B_1$ & $B_2$ & $B_3$ & $B_4$  \\
\hline
 $\left(
\begin{array}{cccc}
\times & \times & 0 &\times \\  \times &0 & \times & \times \\ 0 & \times & \times & \times \\\times & \times & \times & \times
\end{array}
\right)$  & $\left(
\begin{array}{cccc}
\times& 0 & \times &\times \\  0 &\times & \times & \times \\ \times & \times & 0& \times \\\times & \times & \times & \times
\end{array}
\right)$& $\left(
\begin{array}{cccc}
\times & 0 & \times &\times \\  0 & 0 & \times & \times \\ \times & \times & \times & \times \\\times & \times & \times & \times
\end{array}
\right)$  & $\left(
\begin{array}{cccc}
\times & \times& 0 &\times \\  \times & \times & \times & \times \\0 & \times & 0 & \times \\\times & \times & \times & \times
\end{array}
\right)$ \\
\hline $C$& & &  \\
\hline
 $\left(
\begin{array}{cccc}
\times & \times & \times &\times \\ \times & 0 & \times & \times \\ \times & \times & 0 & \times \\\times & \times & \times & \times
\end{array}
\right)$ & & & \\
\hline $D_1$& $D_2$ & & \\
\hline
$\left(
\begin{array}{cccc}
\times &\times & \times &\times \\  \times & 0 & 0 & \times \\ \times & 0 & \times & \times \\\times & \times & \times & \times
\end{array}
\right)$ & $\left(
\begin{array}{cccc}
\times & \times & \times &\times \\  \times & \times &0 & \times \\ \times & 0 &0 & \times \\\times & \times & \times & \times
\end{array}
\right)$& & \\
\hline $E_1$ & $E_2$ & $E_3$ & \\
\hline
$\left(
\begin{array}{cccc}
0 & \times & \times &\times \\ \times & 0 & \times & \times \\ \times & \times & \times & \times \\\times & \times & \times & \times
\end{array}
\right)$& $\left(
\begin{array}{cccc}
0 & \times & \times &\times \\  \times & \times & \times & \times \\ \times & \times & 0 & \times \\\times & \times & \times & \times
\end{array}
\right)$&  $\left(
\begin{array}{cccc}
0 &\times & \times &\times \\ \times & \times & 0& \times \\ \times & 0 & \times & \times \\\times & \times & \times & \times
\end{array}
\right)$& \\
\hline $F_1$& $F_2$ & $F_3$ & \\
\hline
$\left(
\begin{array}{cccc}
\times & 0 & 0 &\times \\  0 & \times & \times & \times \\0 & \times & \times & \times \\\times & \times & \times & \times
\end{array}
\right)$&  $\left(
\begin{array}{cccc}
\times& 0 & \times &\times \\  0 & \times & 0 & \times \\ \times & 0 & \times & \times \\\times & \times & \times & \times
\end{array}
\right)$ & $\left(
\begin{array}{cccc}
\times &\times & 0 &\times \\ \times & \times &0 & \times \\ 0 & 0 & \times & \times \\\times & \times & \times & \times
\end{array}
\right)$&\\
\hline
\end{tabular}
\end{small}
\end{center}
\end{tiny}
\caption{Allowed two zero textures in the 3+1 scenario. The 15 possible
two zero textures of three neutrino case are same as these after omitting 
the 4th row and column.  }
\end{small}
\label{Table:2zero}
\end{table}

\section{Formalism}

We consider the 3+1 mass spectrum.
This can generate two possible mass orderings.
In one case the fourth neutrino is heavier than the
other three and in the other case the fourth neutrino is
lighter than the other three. LSND/MiniBooNE observations
dictate that the mass squared difference of the fourth state with
the three other states is $\sim$ eV$^2$.
However, the scheme in which the fourth state is lower
would be more disfavored from cosmology since there will be
three neutrino states with mass $\sim$ eV
which will contribute to the cosmological
energy density. Therefore, we consider the picture in which the
fourth state is heavier. Then, there are two possibilities shown in
Fig. 1.
 \\
\begin{figure}
 \begin{center}
 \includegraphics[scale=0.5,angle=270]{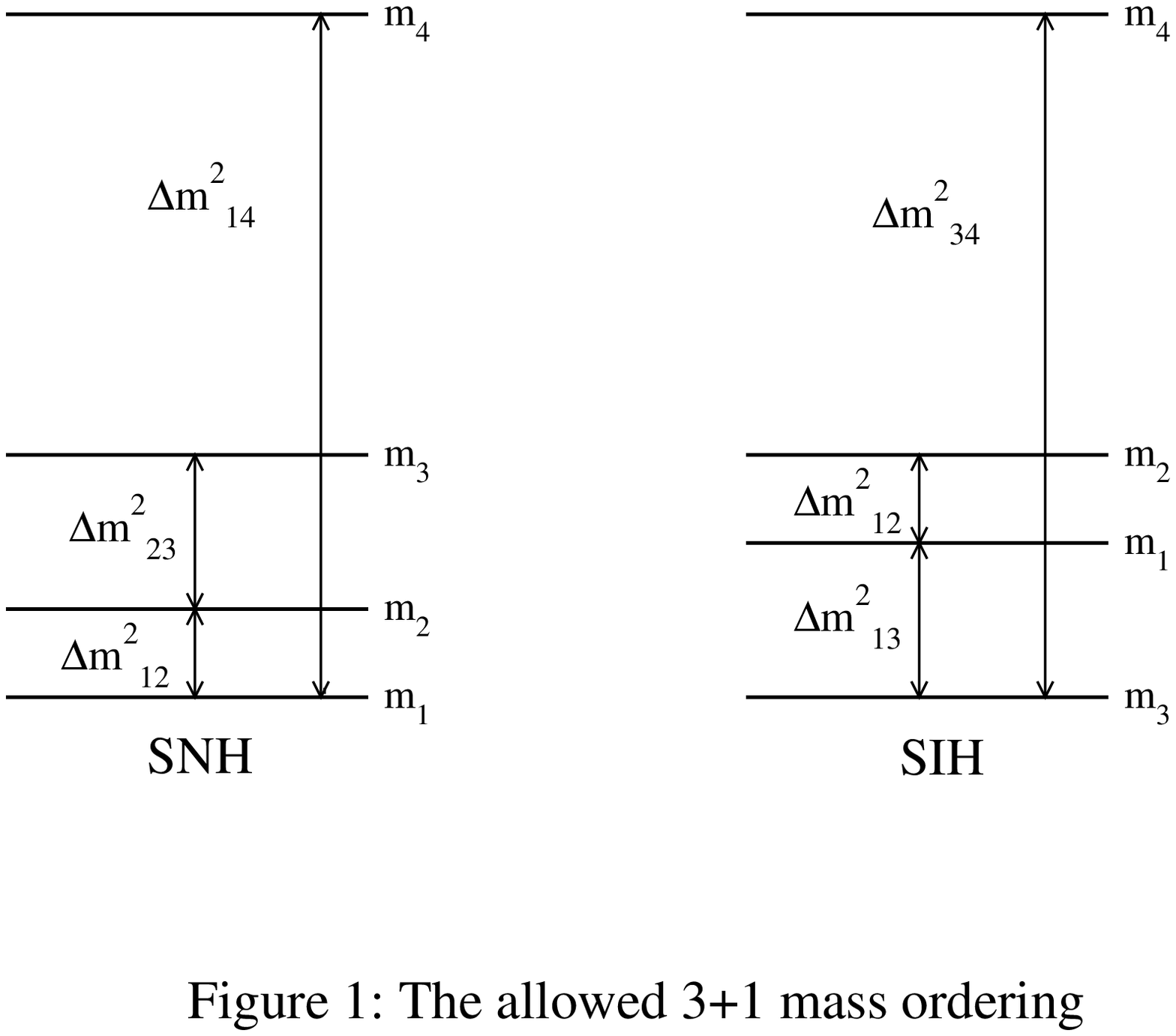}
 \end{center}
\label{fig1}
 \end{figure}

\begin{enumerate}
\item
$m_1 \approx m_2 < m_3 < m_4$ which  corresponds to a normal ordering
among the active neutrinos (SNH).
This gives \\
$m_2=\sqrt{m_1^2+\Delta m_{12}^2}~~ ,
m_3= \sqrt{m_1^2+\Delta m_{12}^2+\Delta m_{23}^2}~~ ,
m_4=\sqrt{m_1^2+\Delta m_{14}^2}.
$

\item
$m_3 < m_2 \approx m_1 < m_4$ corresponding an inverted ordering
among the active neutrinos (SIH) with the masses
\\
$m_1= \sqrt{m_3^2+\Delta m_{13}^2}~~,
m_2=\sqrt{m_3^2+\Delta m_{13}^2+\Delta m_{12}^2}~~,
m_4=\sqrt{m_3^2+\Delta m_{34}^2}.
$
\end{enumerate}
Here, $\Delta m_{ij}^2 = m_j^2 - m_i^2$. 

\begin{table}
\begin{tabular}{lccc}
\hline
\hline
Parameter & Best fit & $1\sigma$ range & $3\sigma$ range \\
\hline
$\Delta m^2_{12}/10^{-5}~\mathrm{eV}^2 $ (NH or IH) & 7.54 & 7.32 -- 7.80 & 6.99 -- 8.18 \\
\hline
$\sin^2 \theta_{12}/10^{-1}$ (NH or IH) & 3.07 & 2.91 -- 3.25  & 2.59 -- 3.59 \\
\hline
$\Delta m^2_{23}/10^{-3}~\mathrm{eV}^2 $ (NH) & 2.43 & 2.33 -- 2.49  & 2.19 -- 2.62 \\
$\Delta m^2_{13}/10^{-3}~\mathrm{eV}^2 $ (IH) & 2.42 & 2.31 -- 2.49  & 2.17 -- 2.61 \\
\hline
$\sin^2 \theta_{13}/10^{-2}$ (NH) & 2.41 & 2.16 -- 2.66 & 1.69 -- 3.13 \\
$\sin^2 \theta_{13}/10^{-2}$ (IH) & 2.44 & 2.19 -- 2.67 & 1.71 -- 3.15 \\
\hline
$\sin^2 \theta_{23}/10^{-1}$ (NH) & 3.86 & 3.65 -- 4.10  &                       3.31 -- 6.37 \\
$\sin^2 \theta_{23}/10^{-1}$ (IH) & 3.92 & 3.70 -- 4.31  & 3.35 -- 6.63 \\
\hline
$ \Delta m_{LSND}^2(\Delta m_{14}^2 or \Delta m_{34}^2) ~\mathrm{eV}^2$ & 0.89 & 0.80 -- 1.00 & 0.6 -- 2 \\
\hline
$ \sin^2\theta_{14} $ & 0.025 & 0.018 -- 0.033 & 0.01 -- 0.05 \\
\hline
$ \sin^2\theta_{24} $ & 0.023 & 0.017 -- 0.037 & 0.005 -- 0.076 \\
\hline
$ \sin^2\theta_{34} $ & -- &  --  & $ < 0.16 $ \\
\hline
\label{Table:parameters} 
\end{tabular}

\begin{center}
\caption{The experimental constraints on neutrino parameters. The three generation constraints are from global analysis in \cite{fogli}. The constraints on
sterile parameters involving the fourth neutrino are from \cite{giunti12}. The constraint on $ \sin^2\theta_{34}$ is obtained from \cite{thomas-talk}.  }
 \end{center}
\end{table}

We assume that the charged lepton mass matrix is diagonal
and the mixing in the neutrino sector is solely responsible for the
leptonic mixing.
In the present case, the neutrino mixing matrix, $V$ can be parametrized in
terms of six mixing angles ($\theta_{13}$,$\theta_{12}$,$\theta_{14}$,$\theta_{23}$,$\theta_{24}$,$\theta_{34}$), three Dirac type CP 
phases ($\delta_{13}$,$\delta_{14}$,$\delta_{24}$) and three Majorana type CP 
phases ($\alpha$,$\beta$,$\gamma$).
The neutrino mass matrix  in flavour basis is given by
\begin{equation}
M_{\nu}=V^*M_{\nu}^{diag}V^{\dagger}
\label{mnu}
\end{equation} where, $M_{\nu}^{diag} = Diag({m_1,m_2,m_3,m_4})$.\\
$V=U.P$ \cite{gr1} with
\begin{equation}
U={R_{34}}\tilde{R_{24}}\tilde{R_{14}}R_{23}\tilde{R_{13}}R_{12}
\end{equation}
where $R_{ij}$ represent rotation in the \textit{ij} generation space, for instance:\\
\begin{center}
$R_{34}$=$\left(
\begin{array}{cccc}
1~ &~0 & 0 & 0 \\  0~ &~ 0 & 1 & 0 \\ 0~ & ~0 & c_{34}& s_{34} \\0 ~& ~0 & -s_{34} & c_{34}
\end{array}
\right)$ , $\tilde{R_{14}}$=$\left(
\begin{array}{cccc}
c_{14}~ & ~0 &~ ~0 &~s_{14}e^{-i \delta_{14}} \\ 0 ~ & ~ 1&~~ 0 & 0 \\ 0 ~& ~0 &~~ 1 & 0 \\-s_{14}e^{i \delta_{14}}  & ~ 0& ~~0 &c_{14}
\end{array}
\right)$ \\
\end{center}
with $s_{ij}=sin \theta_{ij}$ and $c_{ij}=cos \theta_{ij}$. The diagonal phase matrix has the form\\
\begin{center}
$P=Diag(1,e^{-i \alpha/2}, e^{-i {(\beta/2-\delta_{13})}},e^{-i {(\gamma/2-\delta_{14})}})$.  
\end{center}
The best-fit values and the 1$\sigma$ and 3$\sigma$ ranges of the oscillation parameters
in the 3+1 scenario are given in Table 2. 
One can define three mass ratios
\be
x  =  \frac{m_1}{m_2}e^{i\alpha},~~
y =  \frac{m_1}{m_3}e^{i\beta},~~
z  =  \frac{m_4}{m_1}e^{-2i(\gamma/2-\delta_{14}).}
\label{xyz}
\ee
The two zero textures in the neutrino mass matrix give two complex
equations viz.
 \begin{eqnarray}
M_{\nu (ab)}=0, \\ \nonumber M_{\nu (pq)}=0.
\label{mnuzero}
\end{eqnarray}
where \textit{a, b, p} and \textit{q} can take the values $e$, $\mu$, $\tau$ and $s$.
The above eqn.(2.4) can be written as
\begin{equation}
U_{a1} U_{b1} + \frac{1}{x} U_{a2} U_{b2} + \frac{1}{y} U_{a3} U_{b3} e^{2 i \delta_{13}} + z U_{a4} U_{b4}=0,
\label{M11}
\end{equation}
\begin{equation}
U_{p1} U_{q1} + \frac{1}{x} U_{p2} U_{q2} + \frac{1}{y} U_{p3} U_{q3} e^{2 i \delta_{13}} + z U_{p4} U_{q4}=0.
\label{M22}
\end{equation}
Solving eqns. (\ref{M11}) and (\ref{M22}) simultaneously we get the two mass ratios as
\begin{equation}
x=\frac{U_{a3}U_{b3}U_{p2}U_{q2}-U_{a2}U_{b2}U_{p3}U_{q3}}{U_{a1}U_{b1}U_{p3}U_{q3}-U_{a3}U_{b3}U_{p1}U_{q1}+z(U_{a4}U_{b4}U_{p3}U_{q3}-U_{a3}U_{b3}U_{p4}U_{q4})},
\end{equation}
\begin{equation}
y=-\frac{U_{a3}U_{b3}U_{p2}U_{q2}+U_{a2}U_{b2}U_{p3}U_{q3}}{U_{a1}U_{b1}U_{p2}U_{q2}-U_{a2}U_{b2}U_{p1}U_{q1}+z(U_{a4}U_{b4}U_{p2}U_{q2}-U_{a3}U_{b3}U_{p4}U_{q4})}e^{2 i \delta_{13}}.
\end{equation}
The modulus of these quantities gives the magnitudes
$x_m$, $y_m$ while the argument determines
the Majorana phases $\alpha$ and $\beta$.
\be
x_m = \left|x\right|,~~
y_m  =  \left|y\right|
\ee
\be
\alpha=arg\left(x\right),~~
\beta=arg\left(y\right).
\ee
Thus, the number of the free parameters is five, the lowest
mass $m_1$ (NH) or $m_3$ (IH), three Dirac and one Majorana type CP phases.
We can check for the two mass spectra in terms of the magnitude of the 
mass ratios $x_m$, $y_m$ and $z_m = |z|$  as,
\begin{itemize}
\item
SNH which corresponds to  $x_m<1$ , $y_m<1$ and $z_m>1$
\item
SIH which implies
$x_m<1$ , $y_m>1$ and $z_m>1$
\end{itemize}
\vskip 1.0cm

Thus, it is $y_m$ which determines if the hierarchy among the
three light neutrinos is normal or inverted. Note that if
the three light neutrinos are quasi-degenerate then we will have $x_m \approx y_m \approx 1$.
Unlike the three generation case discussed in \cite{shivani,Two-Zero-papers}
the lowest mass can not be determined in the four neutrino analysis
in terms of $x_m$ and $y_m$ since these ratios also depend on $m_1$
through $z$. Thus, we keep the lowest mass as a free parameter.
To find out the allowed two zero textures we adopt the following procedure.\\
We vary the lowest mass  randomly from 0 to  0.5 eV.
All the five mixing angles in Table 2
(apart from $\theta_{34}$) 
and the three mass squared differences
are distributed normally
about the best-fit values with the $1\sigma$ errors as given in Table 2.
The three Dirac and one Majorana type CP phase as well as
the remaining mixing angle $\theta_{34}$ are randomly generated.
Then, we use the above conditions to find out which
mass spectrum is consistent with the
particular texture zero structure under consideration.
We also calculate the three mass squared difference ratios
\begin{eqnarray}
R_\nu=\frac{\Delta m_{21}^2}{|\Delta m_{23}^2|}=\frac{1-x_m^2}{ |(x_m^2/y_m^2)-1|}, \nonumber \\
R_{\nu1}= \frac{|\Delta m_{31}^2|}{\Delta m_{41}^2}=\frac{|1-y_m^2|}{y_m^2(z_m^2-1)}, \nonumber \\
R_{\nu2}=\frac{\Delta m_{21}^2}{\Delta m_{41}^2}=\frac{1-x_m^2}{ x_m^2(z_m^2-1)}.
\end{eqnarray}
The  $3 \sigma$ ranges of these three ratios calculated from the experimental data are
\begin{eqnarray}
R_{\nu}
&=& (0.02-0.04), \nonumber \\
R_{\nu1} &=& (1.98 \times 10^{-3}-3.3 \times 10^{-3}),  \nonumber \\
R_{\nu2} &=& (0.63 \times 10^{-4}-1.023 \times 10^{-4}).
\label{Reqn}
\end{eqnarray}
The allowed textures are selected by checking that they
give the ratios  within the
above range.

\section{Results and Discussions}

\begin{figure}
\begin{center}
\includegraphics[width=0.25\textwidth,angle=270]{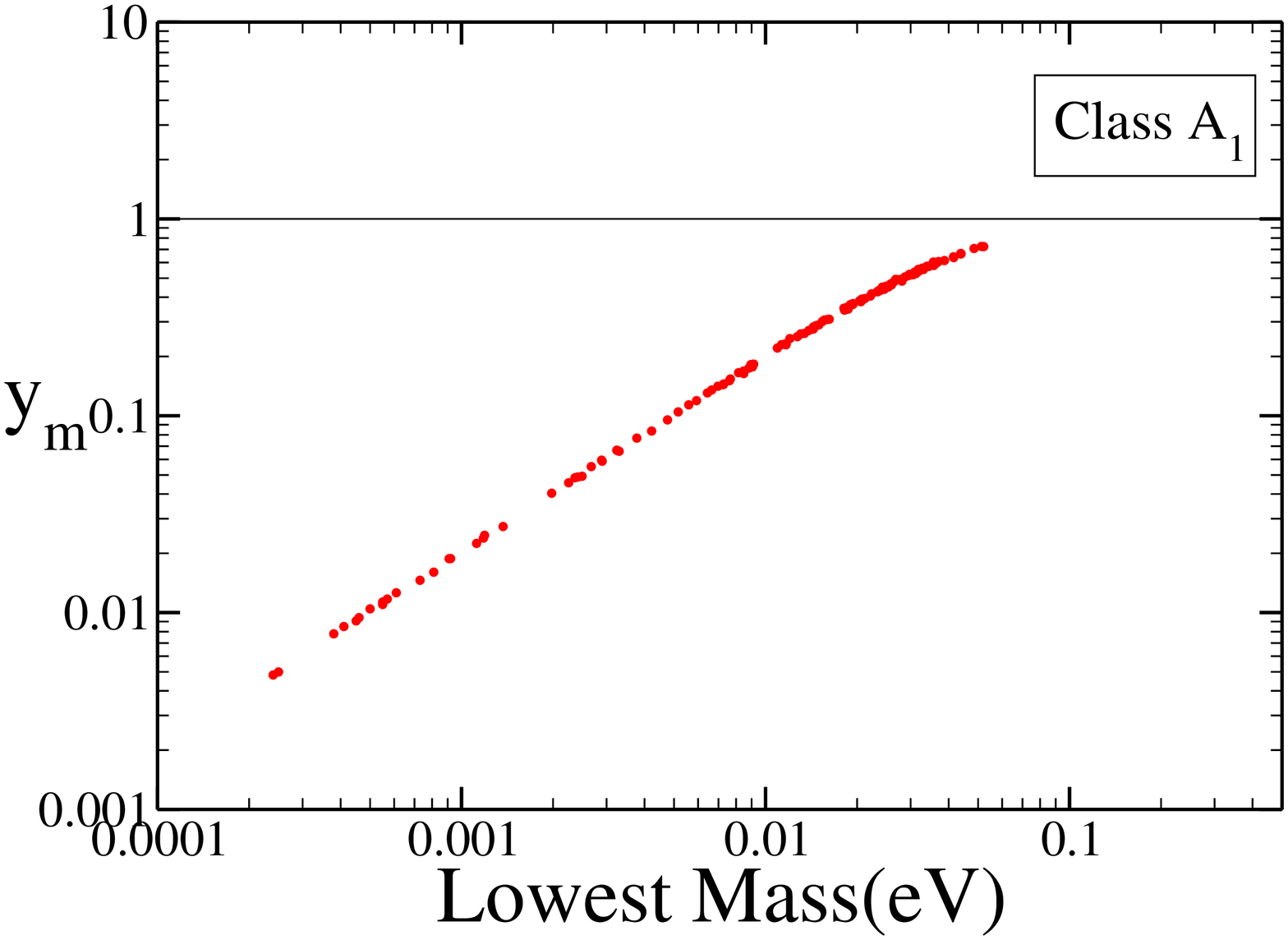}
\includegraphics[width=0.25\textwidth,angle=270]{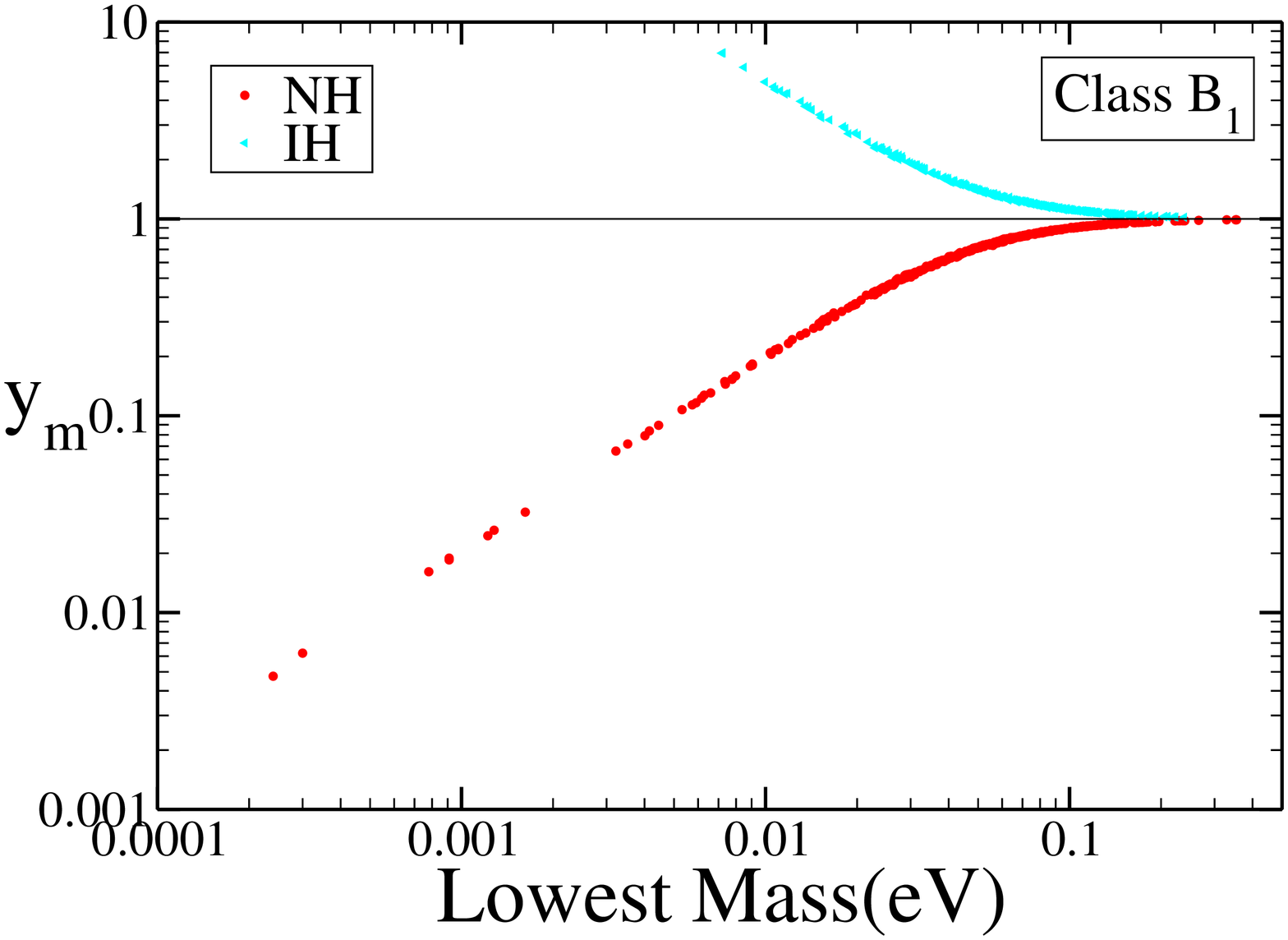}
\includegraphics[width=0.25\textwidth,angle=270]{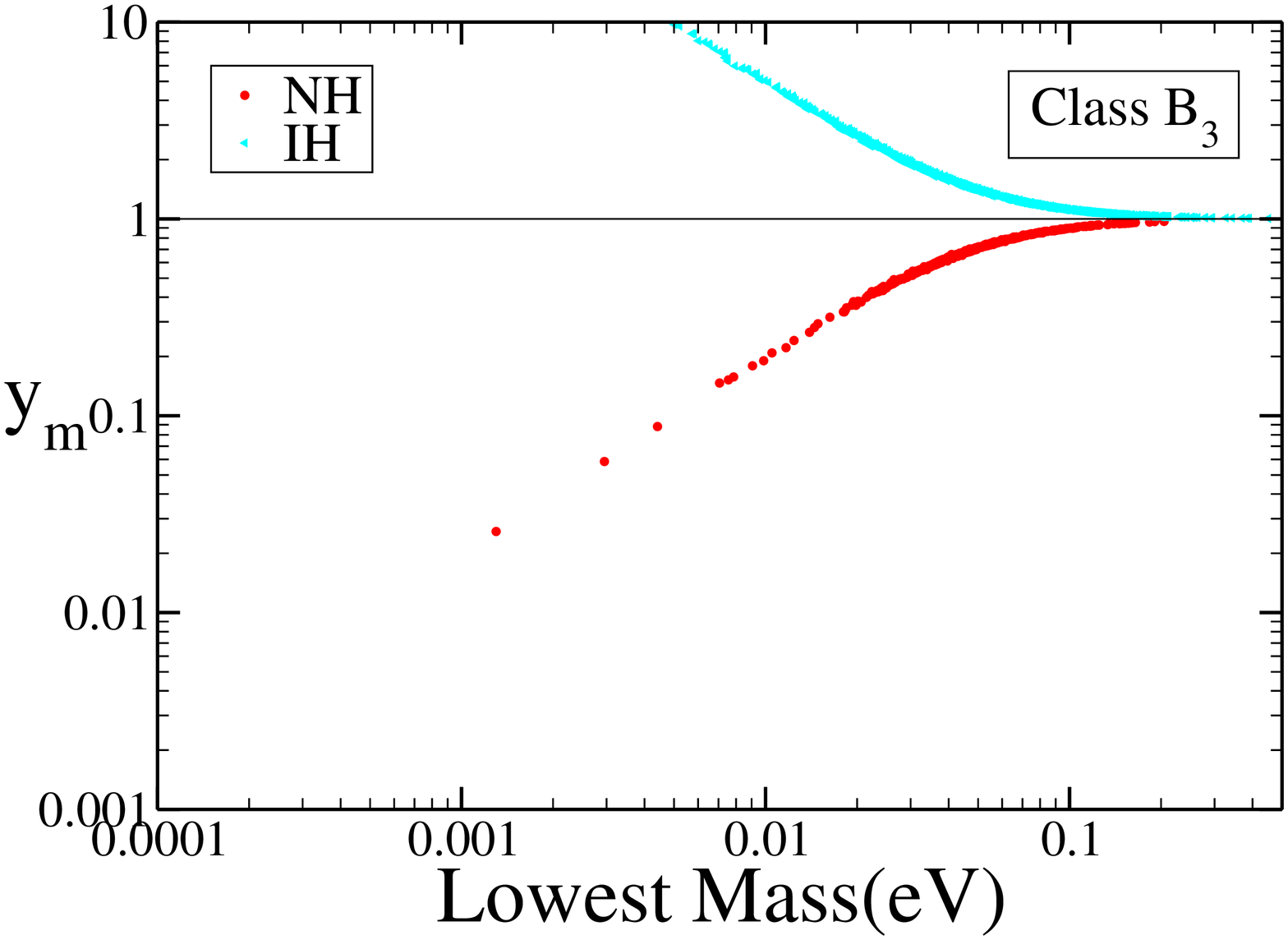} \\
\includegraphics[width=0.25\textwidth,angle=270]{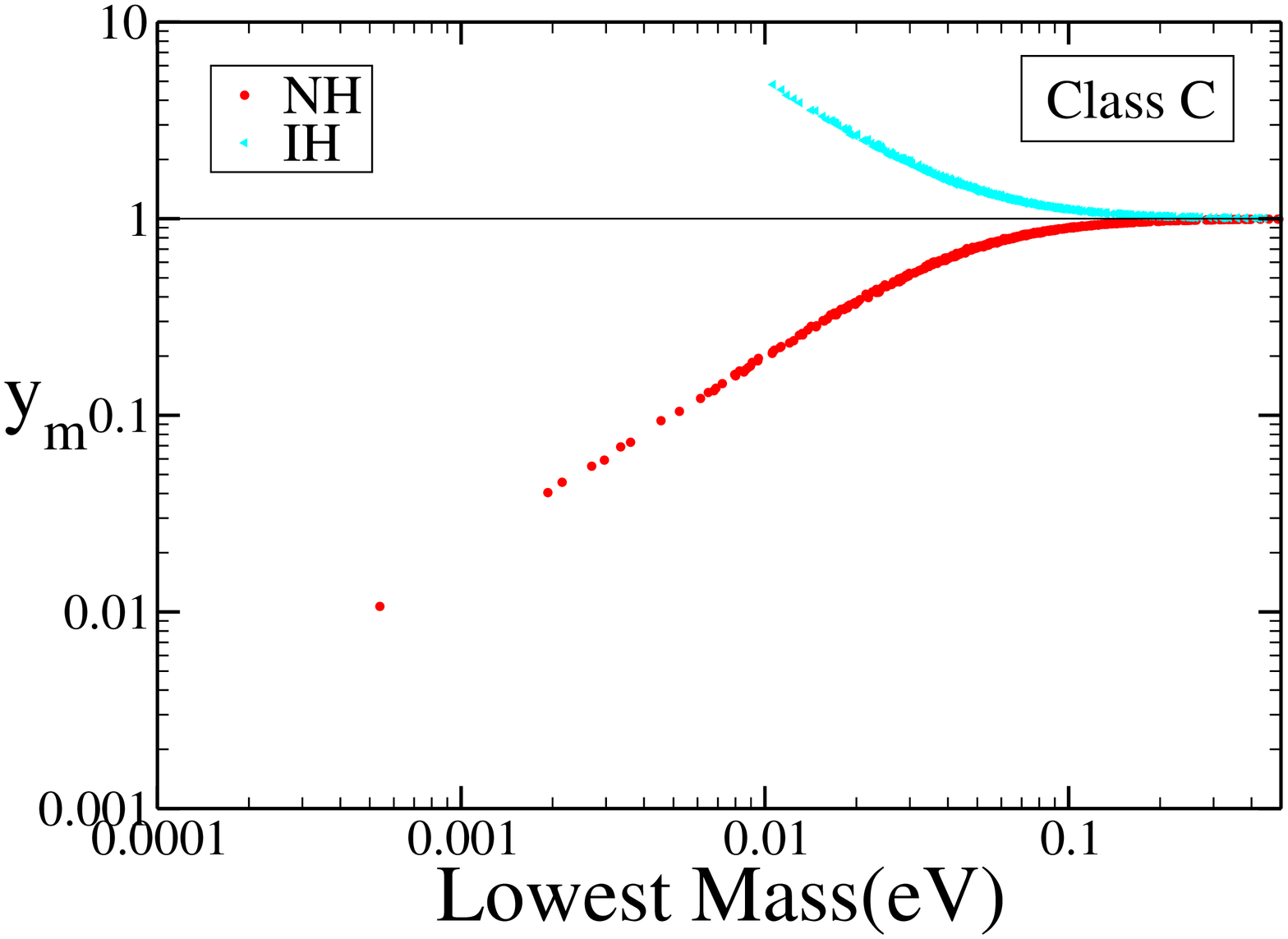}
\includegraphics[width=0.25\textwidth,angle=270]{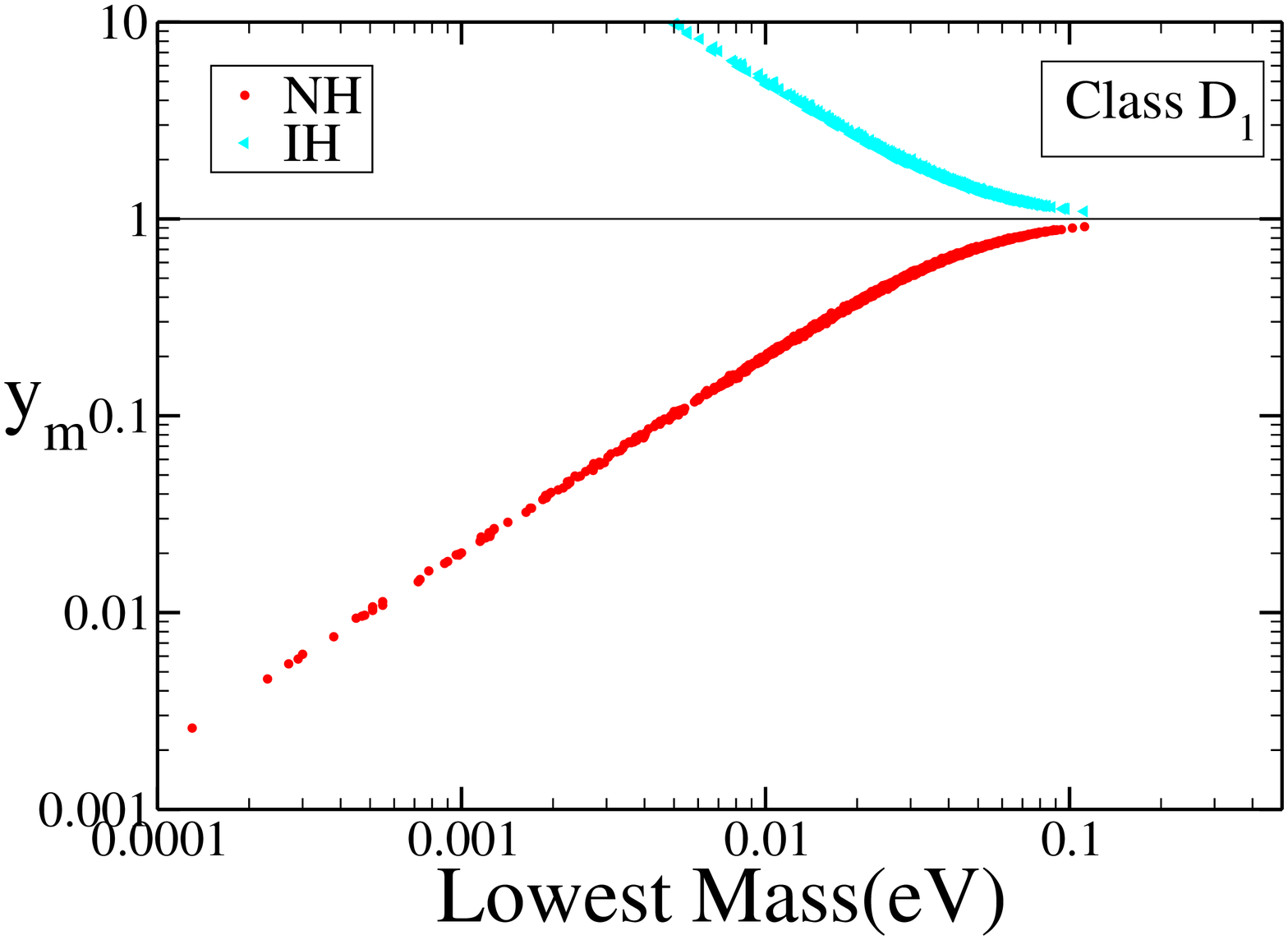}
\includegraphics[width=0.25\textwidth,angle=270]{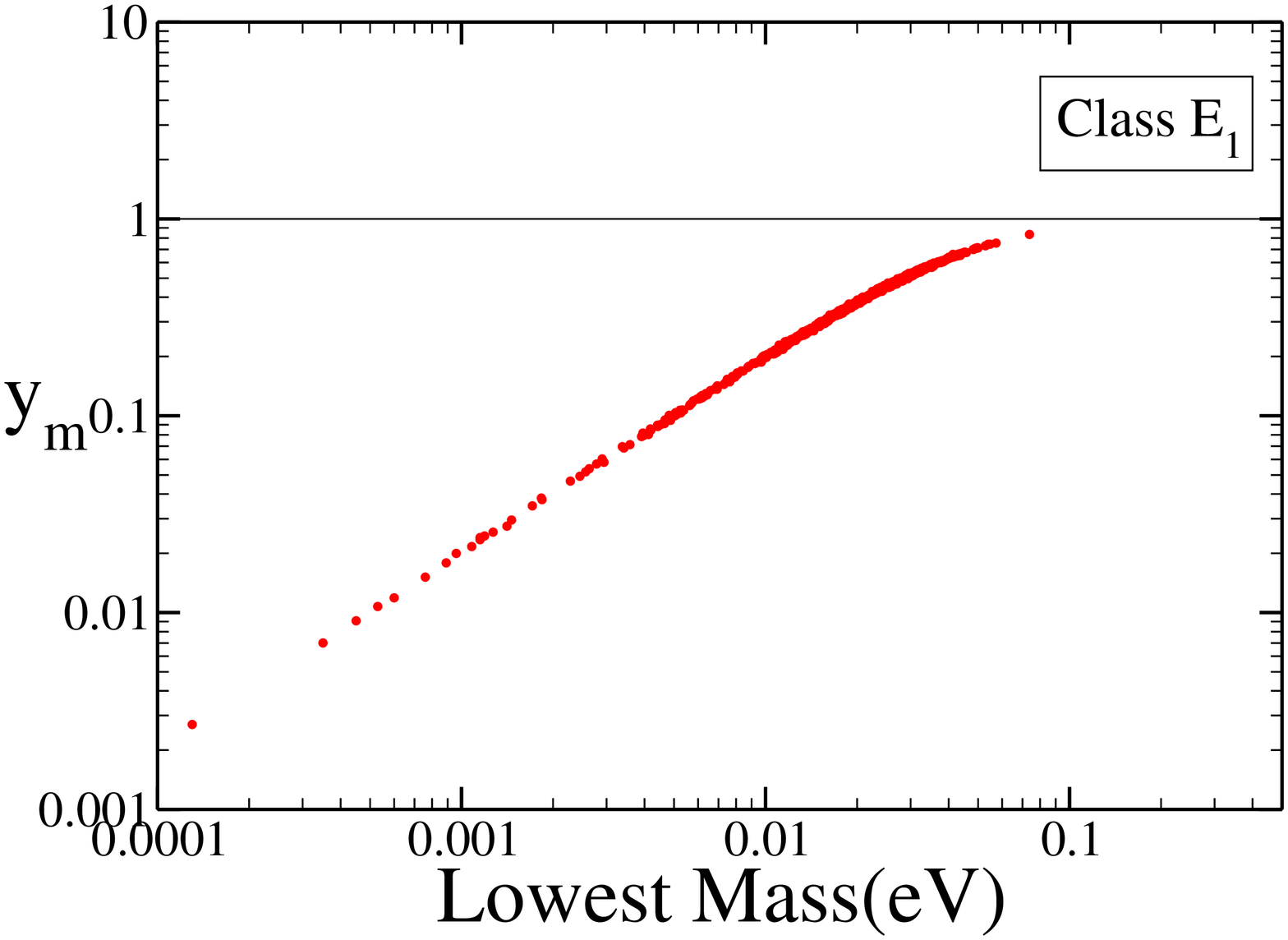} \\
\includegraphics[width=0.25\textwidth,angle=270]{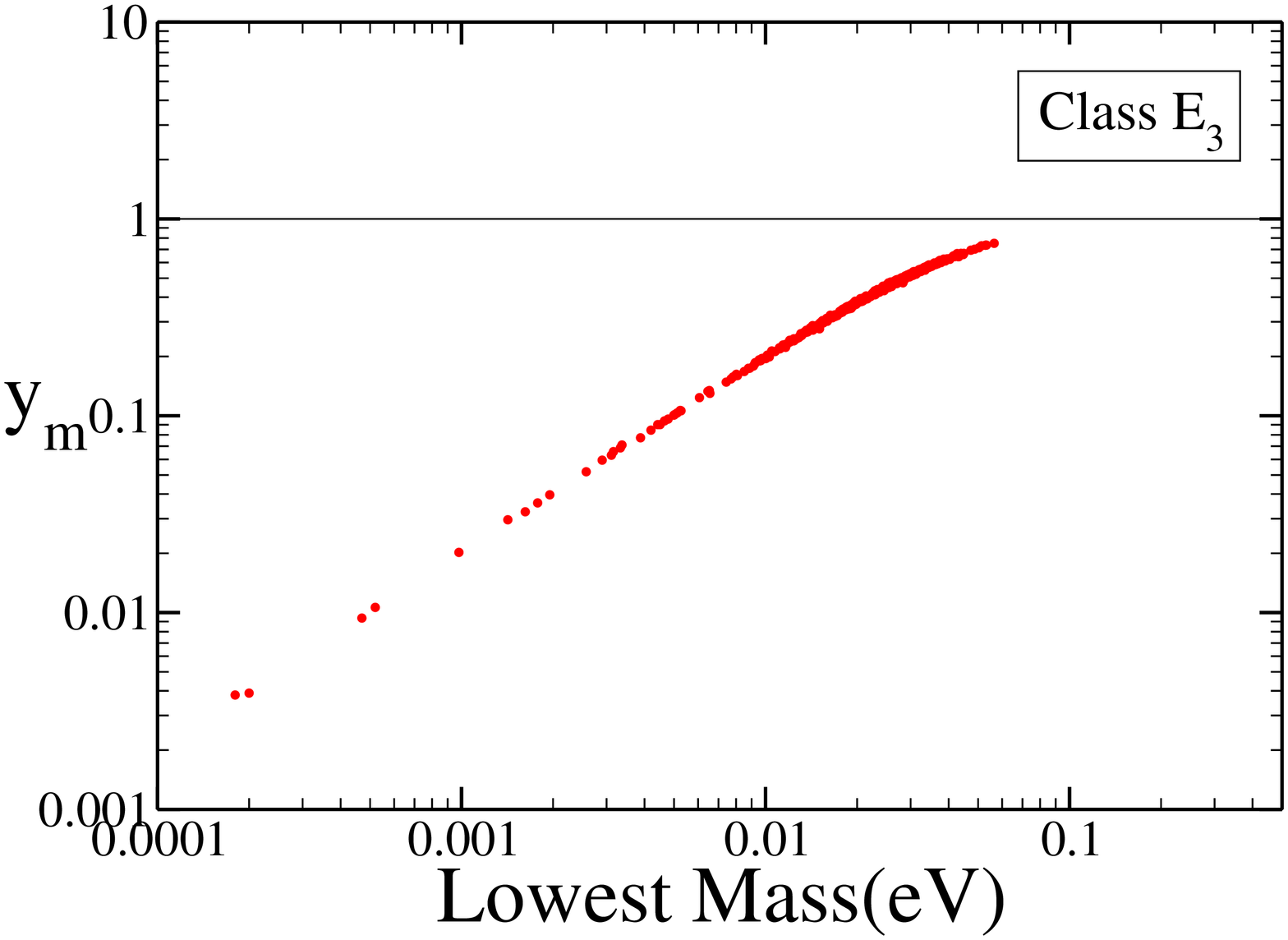}
\includegraphics[width=0.25\textwidth,angle=270]{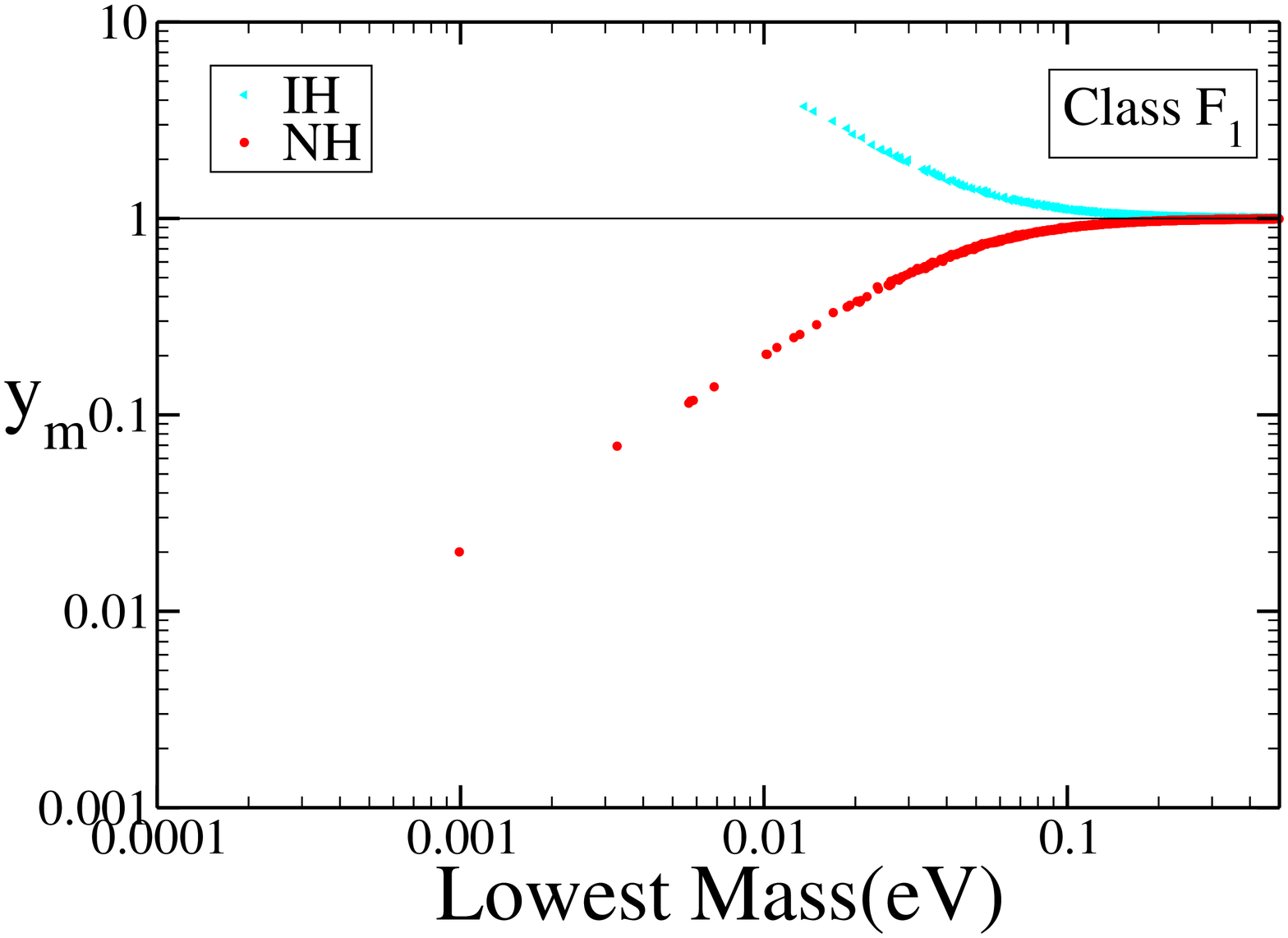}
\includegraphics[width=0.25\textwidth,angle=270]{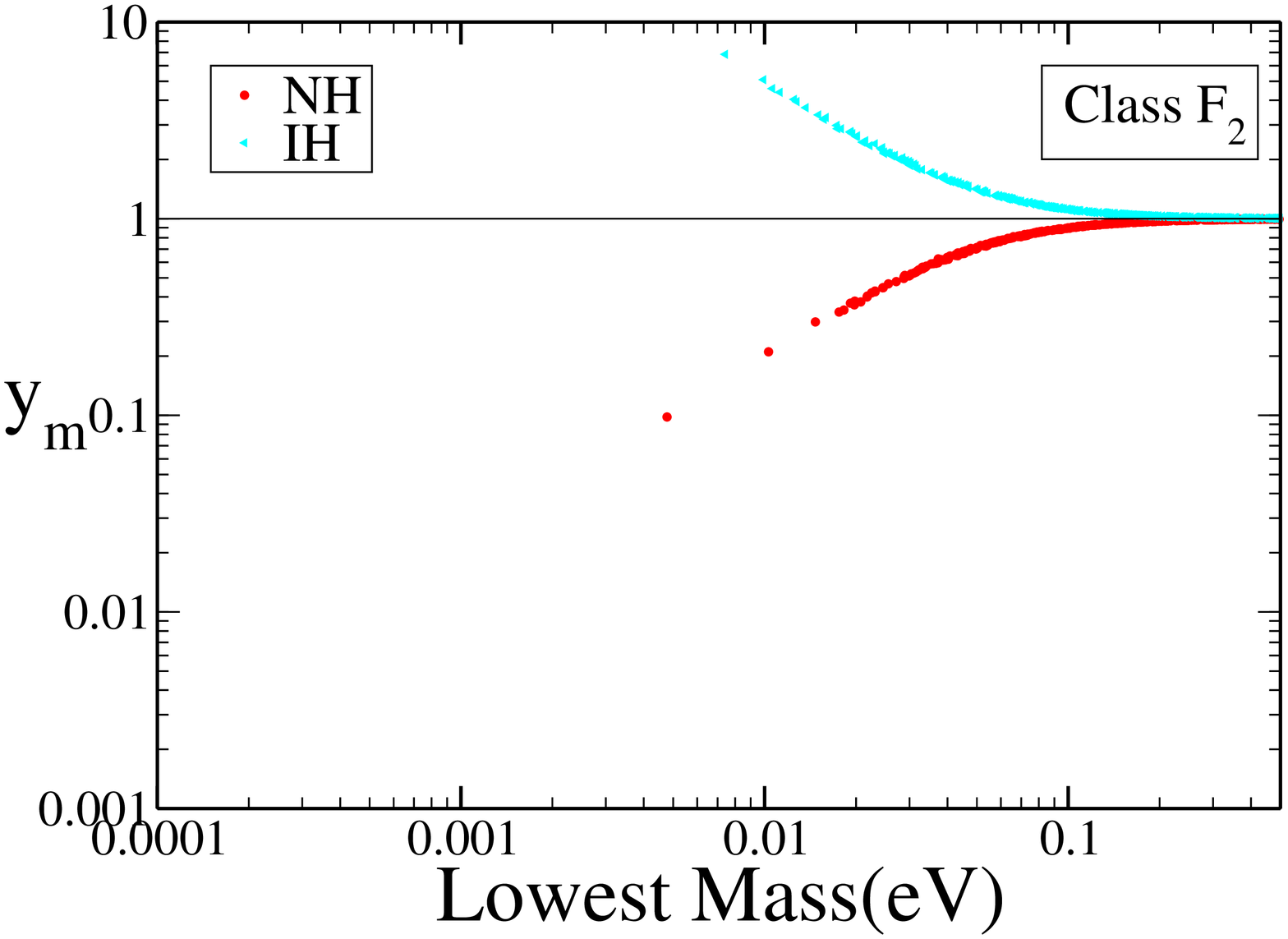}
\end{center}
\begin{center}
Figure 2: The values of $y_m$ as a function of
the lowest mass ($m_1$ or $m_3$) for the 3+1 case
when the known oscillation parameters are varied 
in a Gaussian peaked at their respective best-fit values. 
\end{center}
\end{figure}


\begin{table}
\begin{center}
\begin{tabular}{lccc}
\hline
\hline
Class & 3 generation(Random) & 3 generation(Gaussian) & 3+1 generation (Gaussian)   \\
\hline
 A & NH & NH & NH \\ 
 B & NH, IH, QD & NH($B_1$, $B_3$), IH($B_2$, $B_4$) & NH, IH, QD \\ 
 C & NH, IH, QD & IH & NH, IH, QD \\ 
 D & - &  - & NH, IH \\ 
 E & - & - & NH \\ 
 F & - & - & NH, IH,QD \\
\hline
\end{tabular}
\begin{center}
\caption{The allowed mass spectra in 3 and 3+1  
scenarios. The second and third columns give the results for the two zero neutrino mass matrices for three active neutrinos 
by assuming random and normal distribution of oscillation parameters
respectively. The last column gives the allowed spectrum in the 
for the 3+1 case
assuming normal distribution of parameters. For random distribution of oscillation parameters similar mass spectra get allowed although the parameter space 
is reduced in size. 
See text for details.}
 \end{center}
 \label{2results}
 \end{center}
\end{table}

In this section we present the results of our analysis. 
First we briefly discuss the results that we obtain for the
two zero textures of the $3 \times 3$ mass matrices.
Next we present the results that we obtain for the 3+1 scenario i.e $4 \times4$ mass matrices. 

\subsection{Results for 3 neutrino mass matrix} 

For the 3 neutrino case,  
the lowest mass and the two Majorana phases can be determined 
from the mass ratios. 
Hence, the only unknown parameter 
is the Dirac type CP phase ($\delta_{13}$) which is generated randomly. 
All the other oscillation parameters 
are distributed normally, peaked at the best-fit and taking their 
one sigma error as width. 
We find that all 7 textures which were allowed previously remain so.
However, the textures belonging to A class allow NH whereas for the B class, 
$B_1$ and $B_3$ admit NH and $B_{2}$ and $B_4$  allow IH solutions. 
Class C gets allowed only for IH. 
The D, E and F classes remain disallowed. 
In the 2nd column of 
Table \ref{2results} we summarize 
the results that we obtain for the two zero neutrino mass matrices 
with three active neutrinos using 
normal distribution of the oscillation parameters. 
The results obtained in this case are somewhat different from
that obtained 
using random distribution of oscillation parameters.  
The reason for the difference stems from  
the  different range of values of the atmospheric mixing angle $\theta_{23}$ 
used by these methods. If we assume a Gaussian 
distribution for $\sin^2\theta_{23}$ around its best-fit then there is very 
less probability of
getting the  3$\sigma$ range in the higher octant as these values lie 
near the tail of the Gaussian distribution. This disallow 
$B_2$ and $B_4$ for NH and $B_1$ and $B_3$ for IH \cite{xingnew}.   
Similarly QD solutions for B class requires $\theta_{23} \sim \pi/4$ 
\cite{frampton} and for a normal distribution of $\theta_{23}$ with 
the peak at present best-fit the 3$\sigma$ range extends upto $\sim 44^o$ and there is very little 
probability of getting values close to $\pi/4$.  
Similarly for the C class NH and QD solutions are 
allowed only for $\theta_{23}$ values 
close to $\pi/4$ and hence is not admissible when Gaussian distribution of 
oscillation parameters about the best-fit value is assumed. 

\subsection{Results for 3+1 scenario} 
 
Adding one sterile neutrino,
there exist in total forty five texture structures of the neutrino
mass matrix which can have two zeros.
\begin{enumerate}

\item [(i)] Among these the 9 cases with $m_{ss} = 0$ are
disallowed as the mass matrix element
$m_{ss}$ contains the term $m_4U_{s4}^2$ which is large
from the current data and suppresses the other terms. Hence, $m_{ss}$ cannot vanish.

\item
[(ii)] There are 21 cases where one has at least one zero involving
the mass matrix element of the sterile part i.e  $m_{ks} =0$
where $k=e,\mu,\tau$.
This element is of the form,
\bea
m_{ks}  & = &
m_1U_{k1}U_{s1}+m_2U_{k2}U_{s2}e^{-i\alpha} \\
\nonumber
& + &
m_3 U_{k3}U_{s3}e^{i(2\delta_{13}-\beta)}+m_4 U_{k4}U_{s4}e^{i(2\delta_{14}-\gamma)}.
\label{mks}
\eea
The last term in this expression contains the product $m_4 U_{s4}$ which is
quite large  as compared to first three terms and thus, there can
be no cancellations. Thus, neutrino mass matrices with one of the
zeros in fourth row or column are not viable.

\item
[(iii)] The remaining cases are the 15 two zero cases
for which none of the sterile components are zero. Thus, these also belong to the two zero textures of the three generation
mass matrix.
A general element in this category can be expressed as,
\bea
m_{kl}  & = &
m_1U_{k1}U_{l1}+m_2U_{k2}U_{l2}e^{-i\alpha} \\
\nonumber
& + &
m_3 U_{k3}U_{l3}e^{i(2\delta_{13}-\beta)}+m_4 U_{k4}U_{l4}e^{i(2\delta_{14}-\gamma)}.
\label{mkl}
\eea
here, $k,l=e,\mu,\tau$.
We find all these 15 textures, presented in Table \ref{Table:2zero} 
get  allowed with the inclusion  of the sterile neutrino.
This can be attributed to additional cancellations that the 
last term in eq. 3.2 induces. 
Table 3 displays the nature of the mass spectra that are 
admissible in the allowed textures. 


In Fig.2 we present the values of $y_m$ vs the lowest mass
for textures $A_1$, $B_1$,
$B_3$, $C$, $D_1$, $E_1$, $E_3$, $F_1$ and $F_2$.
This figure shows that for textures belonging to the A and E classes
$y_m$ remains $<$ 1. Thus, these classes admit only NH solutions. The textures belonging to the D class allow NH and IH 
while the B,C,and F classes allow NH, IH and QD mass spectra.

The textures $A_1-A_2$, $B_1-B_2$, $B_3-B_4$, $D_1-D_2$,
$E_1-E_2$ and $F_2-F_3$ are related by $P_{\mu\tau}$ symmetry where
for the four neutrino framework  $P_{\mu\tau}$ can be expressed as,
\begin{center}
$
P_{\mu\tau}=\left(
\begin{array}{cccc}
 1& 0 & 0 & 0 \\ 0& 0 &1& 0 \\ 0& 1 &0 & 0\\ 0 & 0 & 0 &1
\end{array}
\right)$
\end{center}
in such a way that
\begin{center}$A_2=P_{\mu\tau}^T A_1 P_{\mu\tau} . $ \end{center}

Note that for 3 generation case the angle $\theta_{23}$ in the partner
textures linked by $\mu-\tau$ symmetry was related as
$\bar{\theta_{23}}=(\frac{\pi}{2} - \theta_{23})$.
However, for the 3+1 case no such simple relations are obtained for the mixing
angle $\theta_{23}$. The angles $\theta_{24}$ and $\theta_{34}$ in the two
textures related by $\mu-\tau$ symmetry are also different.
For this case, the mixing angles
for two textures linked  by
$P_{\mu \tau}$ symmetry
are related as
\begin{equation}
\bar{\theta_{12}} = \theta_{12},
~~~ \bar{\theta_{13}} = \theta_{13},
~~~ \bar{\theta_{14}} = \theta_{14},
\end{equation}
\begin{equation}
\sin{\bar\theta_{24}} =  \sin\theta_{34} \cos{\theta_{24}}, 
\end{equation}
\begin{equation}
\sin {\bar\theta_{23}} 
=\frac{\cos{\theta_{23}}\cos{\theta_{34}}-\sin{\theta_{23}}\sin{\theta_{34}}\sin{\theta_{24}}}{\sqrt{1-\cos{\theta_{24}^2}\sin{\theta_{34}^2}}},
\end{equation}
\begin{equation}
\sin {\bar\theta_{34}}
=\frac{\sin{\theta_{24}}}{\sqrt{1-\cos{\theta_{24}^2}\sin{\theta_{34}^2}}}.
\end{equation}


\end{enumerate}

The texture zero conditions together with the constraints imposed by
the experimental data allow us to obtain correlations between various
parameters specially the mixing angles of the 4$^{\mathrm th}$ neutrino
with the other three for the A and E classes. For the B, C, D and F classes one gets constraints
on the effective mass governing $0\nu\beta\beta$.

In order to gain some analytic insight into the results
it is important to understand
the mass scales involved in the problem. The solar mass scale is 
$\sqrt{\Delta m^2_{21}} \approx 0.009$ eV
whereas the atmospheric mass scale is $\sqrt{\Delta m^2_{31}} \approx 0.05$ eV.
Normal hierarchy among the active neutrinos implies
$m_1 << m_2 << m_3$ corresponding to $m_1 \lsim 0.009$ eV.
It is also possible that $m_1 \approx m_2 << m_3$ implying 
$m_1 \sim$ 0.009 eV - 0.1 eV. We call this partial normal
hierarchy. IH corresponds to $m_3 << m_1 \approx m_2$.
If on the other hand $m_1 > 0.1$ eV then
$m_1 \approx m_2 \approx m_3$ which corresponds to quasi-degenerate neutrinos.

\begin{itemize}

\item \underline{\textbf{$A$ and $E$ Class}}

 \begin{figure}
 \begin{center}
 \includegraphics[width=0.25\textwidth,angle=270]{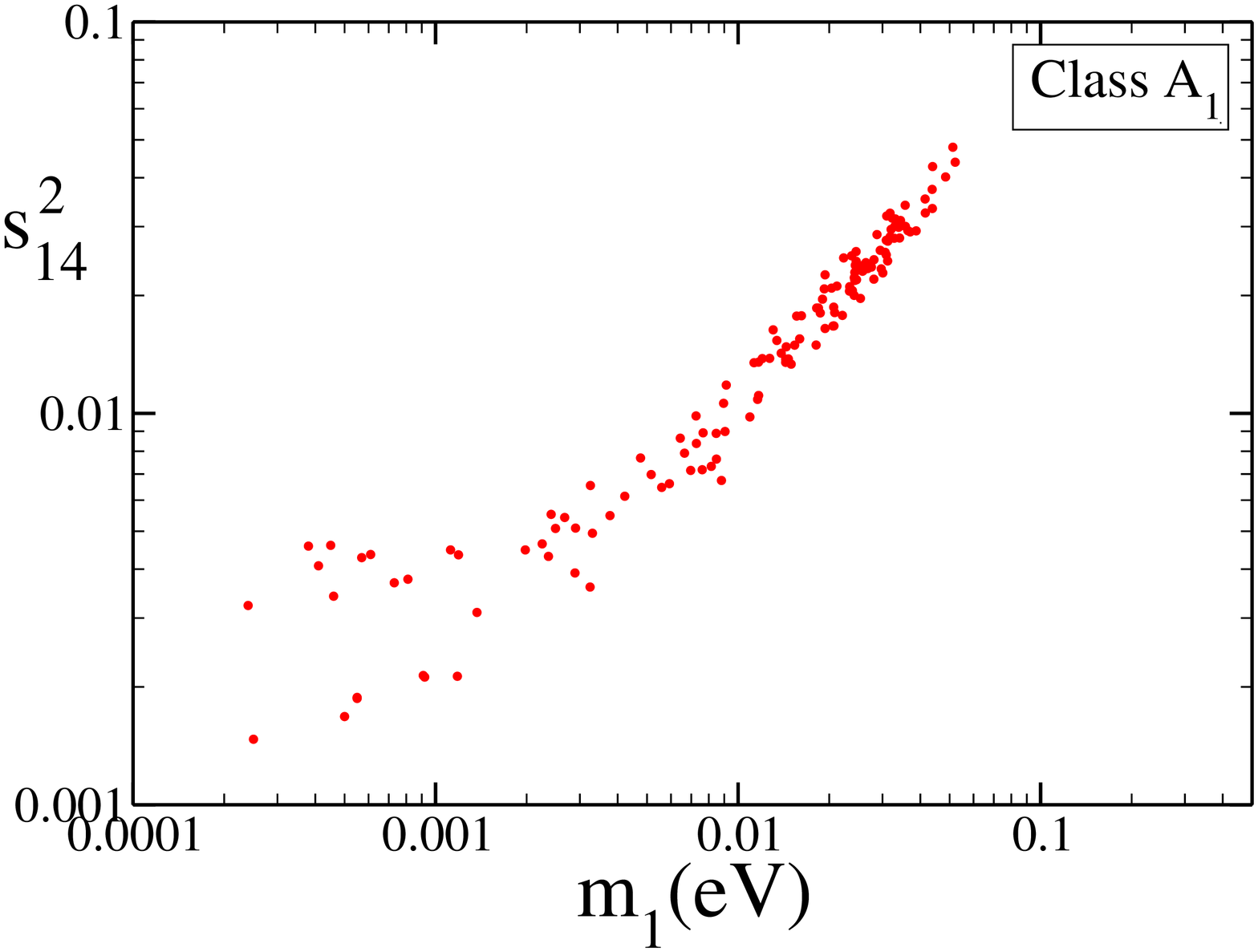}
\includegraphics[width=0.25\textwidth,angle=270]{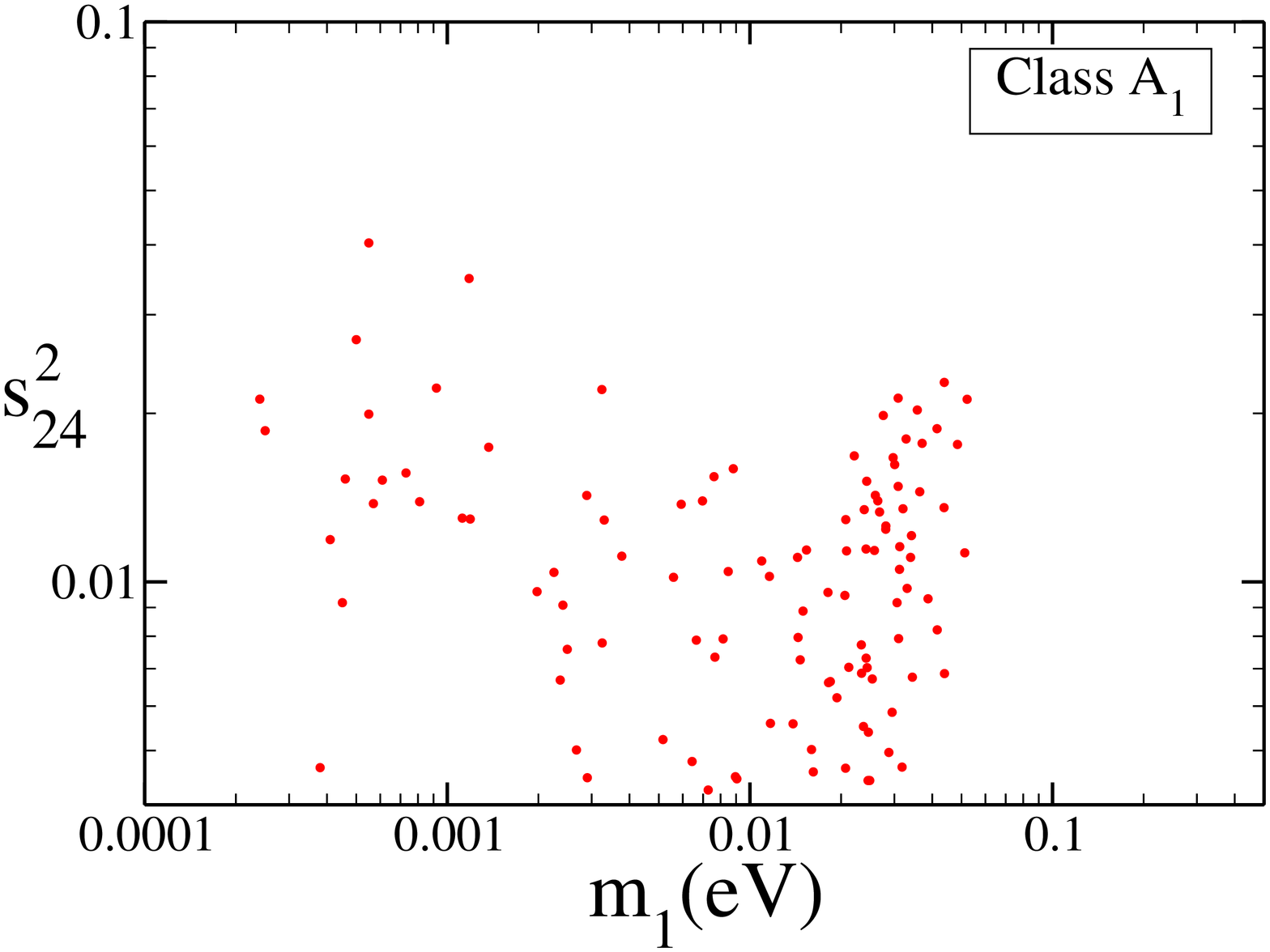}
\includegraphics[width=0.25\textwidth,angle=270]{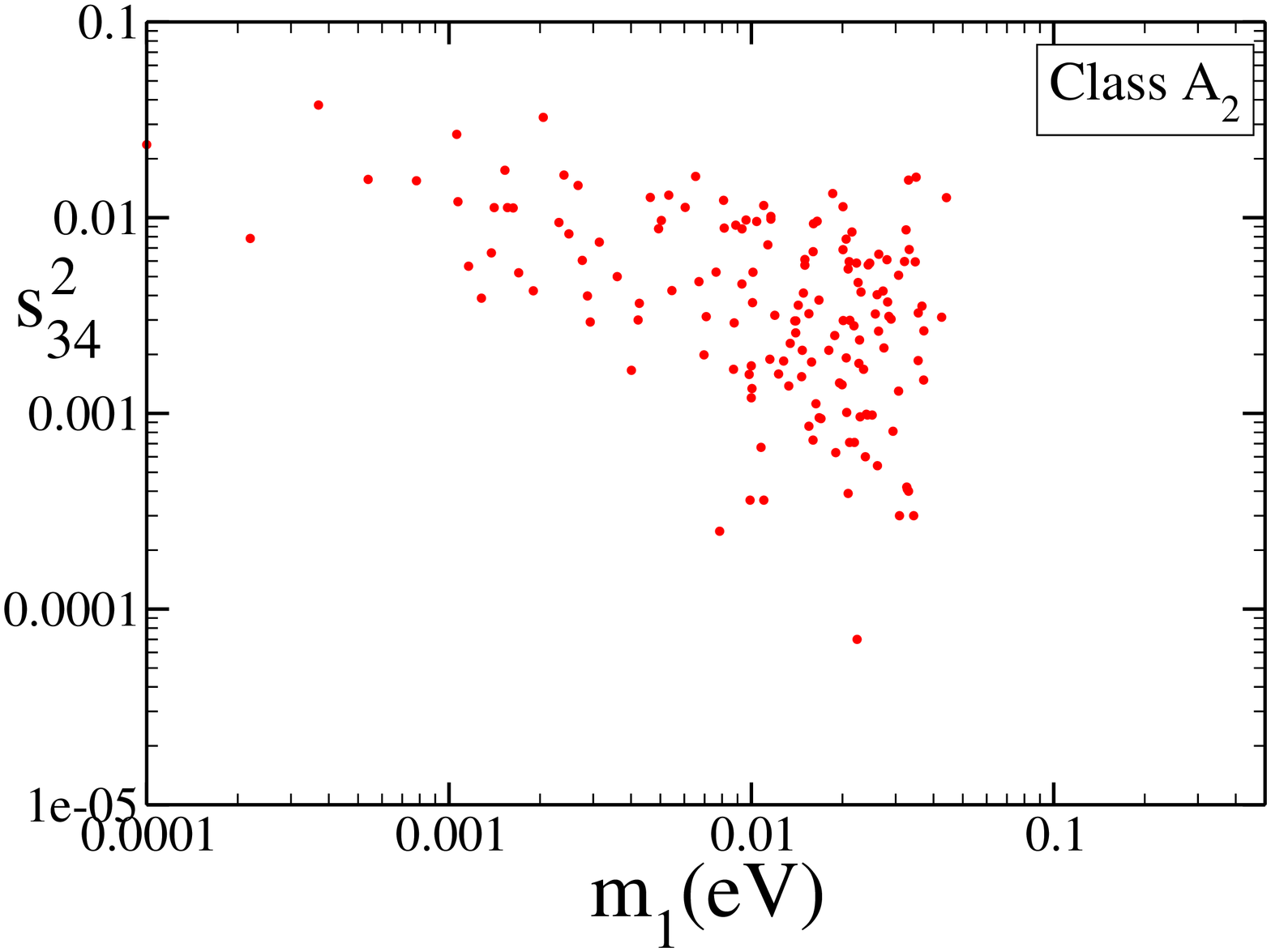} \\
\includegraphics[width=0.25\textwidth,angle=270]{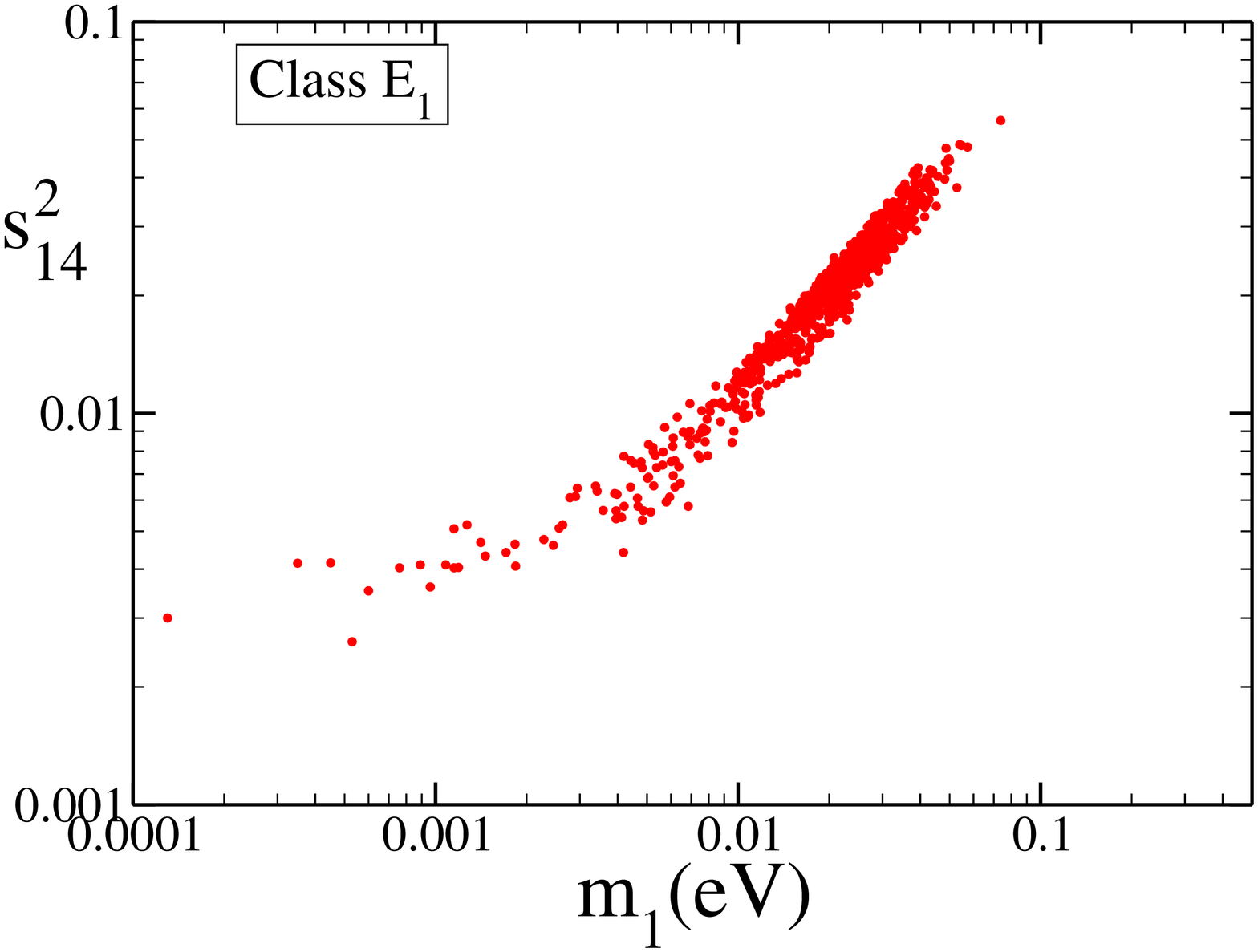}
\includegraphics[width=0.25\textwidth,angle=270]{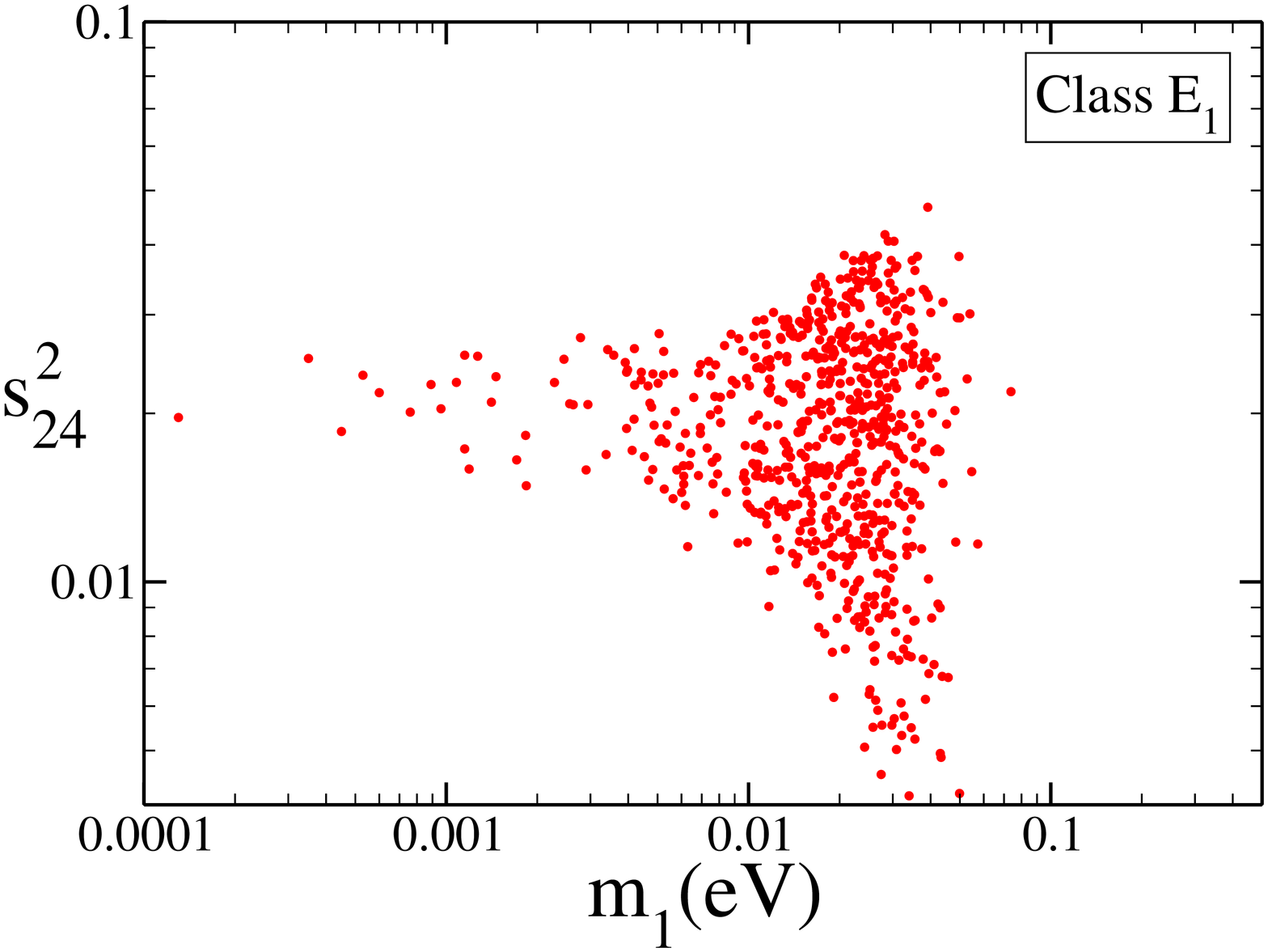}
\includegraphics[width=0.25\textwidth,angle=270]{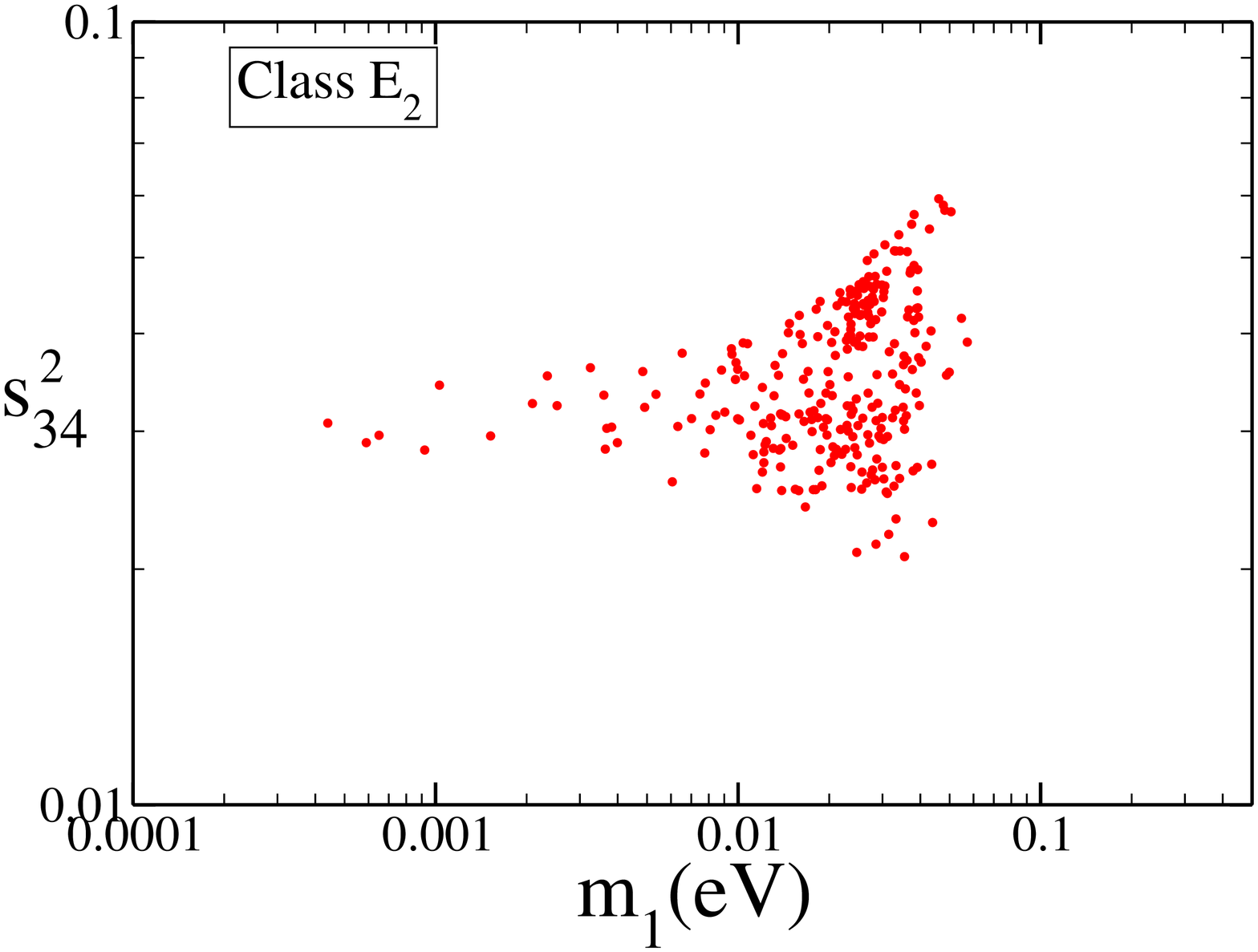}
 \end{center}
 \begin{center}
 Figure 3: Correlation plots for $A$ and $E$ class.
 \end{center}
 \label{Aclass}
 \end{figure}

For these classes we find $y_m$ to be mainly in the range $>$ 0.0001 eV
extending up to $\sim 0.1$ eV.
Thus, these classes allow normal hierarchy (full or partial)
among the 3 active neutrinos.
These classes are characterized by the condition   $m_{ee} =0$.
 $m_{ee}$ for the four neutrino framework can be expressed as,
\bea
m_{ee} & = & c_{12}^2 c_{13}^2 c_{14}^2 m_1 +
c_{13}^2 c_{14}^2 e^{-i \alpha } {m_2} s_{12}^2
+ c_{14}^2
e^{- i \beta } m_3 s_{13}^2 + e^{- i \gamma } {m_4} s_{14}^2.
\label{mee1}
\eea
For smaller values of $m_1$ and NH
the dominant contribution to the magnitude of the above term
is expected to  come from the last term
$s_{14}^2 \sqrt{\Delta m^2_{41}} \sim 0.022$. Therefore, very small values of $m_1$ is less likely to give
$m_{ee}=0$ for normal hierarchy.
However, we get some allowed points in the small $m_1$ regime which
implies smaller values of $s_{14}^2$.
$m_{ee}$ can be approximated in the small $m_1$ limit as,
\be
m_{ee} \approx  e^{-i \alpha } {m_2} s_{12}^2
+
e^{- i \beta } m_3 s_{13}^2 + e^{- i \gamma } {m_4} s_{14}^2.
\ee
The maximum magnitude of the first two terms  is $\sim 0.003$.
Then using typical values of $m_4$ ($\sim 0.9$ eV) from the 3$\sigma$ range,
we obtain
$s_{14}^2 \sim (0.003- 0.004$) in the small $m_1$ limit.
This is true for all the textures in the A and E class.
For the $A_1$ class
we also simultaneously need $m_{e \mu}=0$.
In the small $m_1$ limit approximate expression for $m_{e \mu}$ is
\bea
m_{e \mu} & \approx &
e^{i(\delta _{14}-\delta _{24}- \gamma)}
s_{14} s_{24} m_4 +
e^{i(\delta _{13}-\beta)}  s_{13} s_{23} m_3
+ e^{-i\alpha} c_{12} c_{23}
s_{12} m_2 ,
\label{memu}
\eea
and the first term i.e $m_4 s_{14} s_{24} \sim (0.05 - 0.06) s_{24} $. 
While the other terms are of the order (0.006 - 0.007) which implies
$s_{24}^2 \sim (0.01 - 0.02)$. This is reflected
in the first and second panels of Fig.3 where the correlation of
$s_{14}^2$ and $s_{24}^2$  with $m_1$ is depicted.
As $m_1$ increases the contribution from the
first three terms in $m_{ee}$ increases  and $s_{14}^2$ becomes larger
for cancellation to occur. For $m_{e \mu}$,
this increase in $s_{14}$ helps to achieve cancellation for higher
values of $m_1$ and  therefore $s_{24}^2$ stays almost the same.
Similar argument also apply to the $E_1$ class which has $m_{\mu \mu}=0$.

For $A_{2}$ class, in addition we have $m_{e \tau}=0$.
In the limit of small $m_1$, $m_{e \tau}$ can be approximated as,
\be
m_{e \tau} \approx
e^{i(\delta _{14}- \delta _{24}-\gamma)}
s_{14} s_{34} m_4  +  e^{i(\delta_{13} - \beta)} s_{13}^2
m_3 - m_2 e^{-i \alpha}s_{12} c_{12} s_{23}.
\ee
As discussed earlier $m_{ee}=0$ implies small $s_{14}^2 \sim (0.002 - 0.005$) in
the limit of small $m_1$. Thus, the contribution from the $m_4$ term is $\sim  (0.04 - 0.06) s_{34}$. The typical contribution from the last two terms is $\sim 0.008$. 
This implies $s_{34}^2$ to be in the range (0.02 - 0.04) for smaller values of $m_1$. This is reflected in the third panels of 
Fig.3  where we have plotted the correlation of $s_{34}^2$ with $m_1$.
Since with increasing $m_1$, $s_{14}$ increases
to make $m_{ee}=0$,
$s_{34}^2$
does not increase further.
Similar bounds on $s_{34}^2$ are also obtained for $E_2$ class.

As one approaches the QD regime then the terms containing
the active neutrino masses starts contributing more.
So for higher values of $m_1$ complete cancellation leading to $m_{ee}=0$
even at the highest value of $s_{14}^2$ is not possible. This feature restricts $m_1$ to be $<$ 0.1 eV in A class.

Textures belonging to A and  E class contains $m_{ee}=0$ which is not possible
for IH in the 3 generation case since the solar mixing angle is not
maximal. In the 3+1 scenario a cancellation leading to
$m_{ee} = 0 $ is possible for IH but it requires values of $s_{14}^2$
in the higher side.   
It also contains a strong correlation in the Majorana phases. But in the other mass elements these
constraints are not satisfied simultaneously and as a result the textures that
contain $ m_{ee} =0 $ do not admit inverted hierarchical mass spectrum.

Since $m_{ee} =0$, 
the effective mass (\textit{$m_{eff}$} = $|m_{ee}|$) governing 
the neutrinoless double beta decay ($0\nu\beta\beta$)
is vanishing for these classes.


\begin{figure}
\includegraphics[width=0.25\textwidth,angle=270]{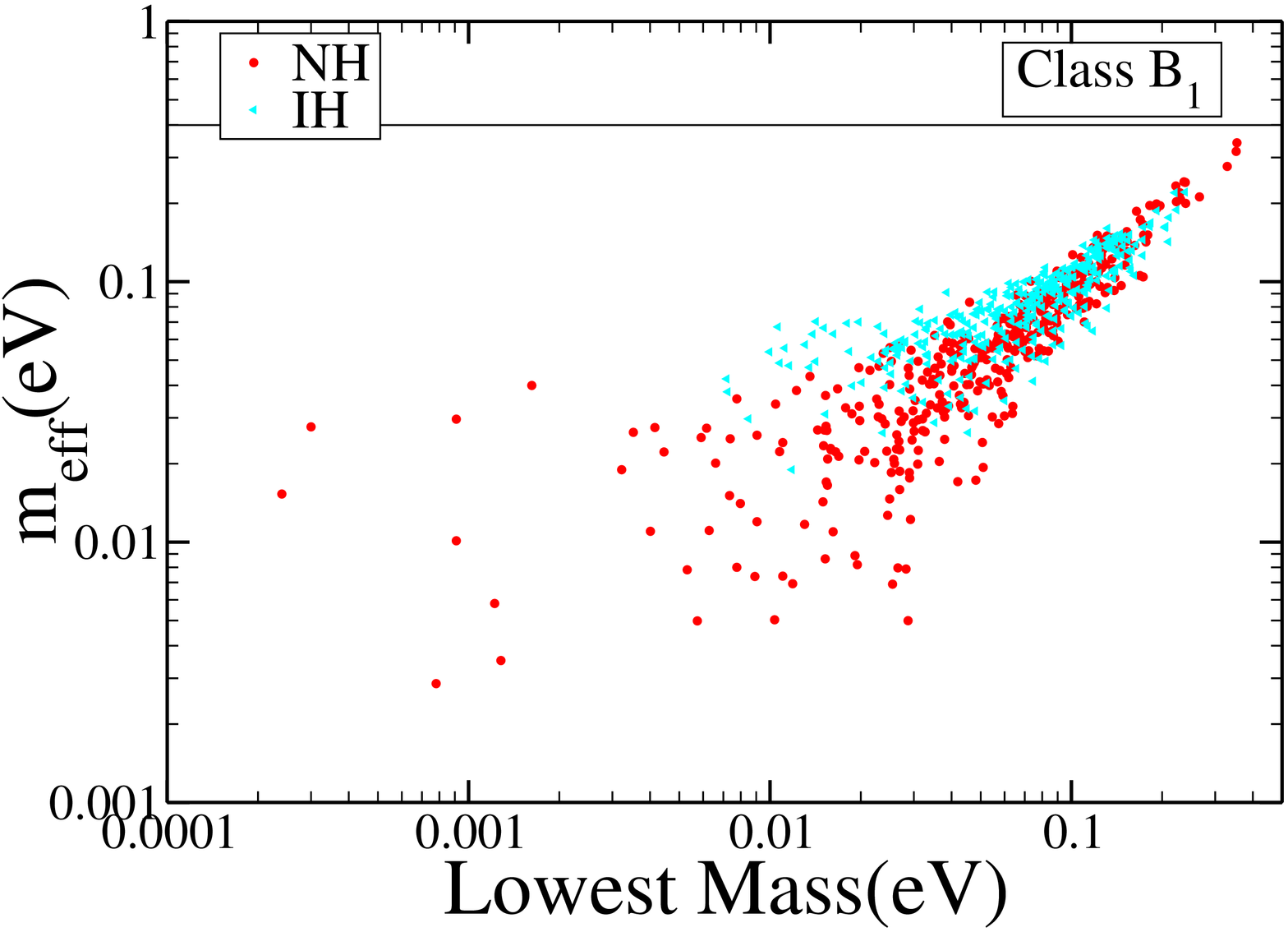}
\includegraphics[width=0.25\textwidth,angle=270]{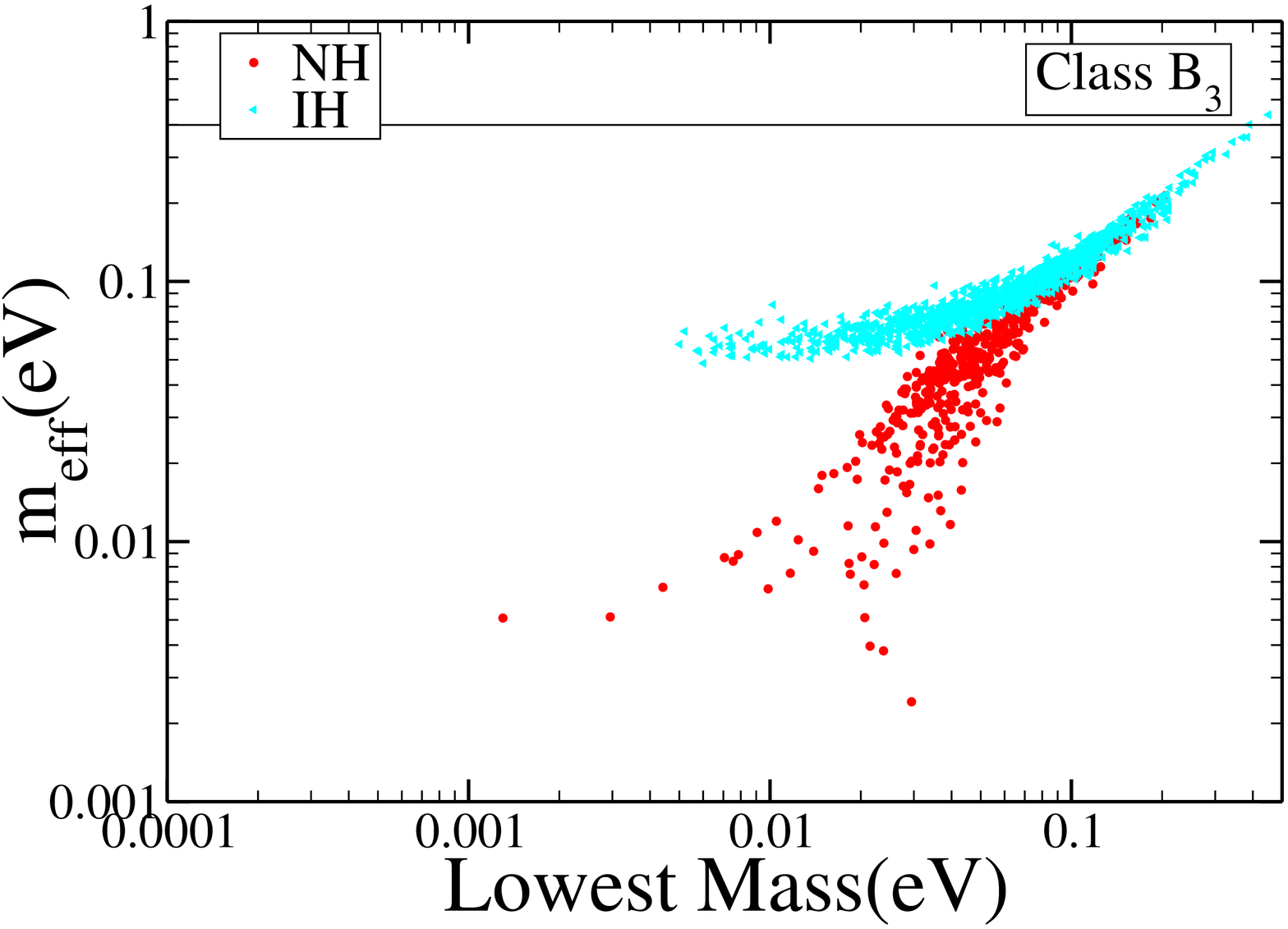}
\includegraphics[width=0.25\textwidth,angle=270]{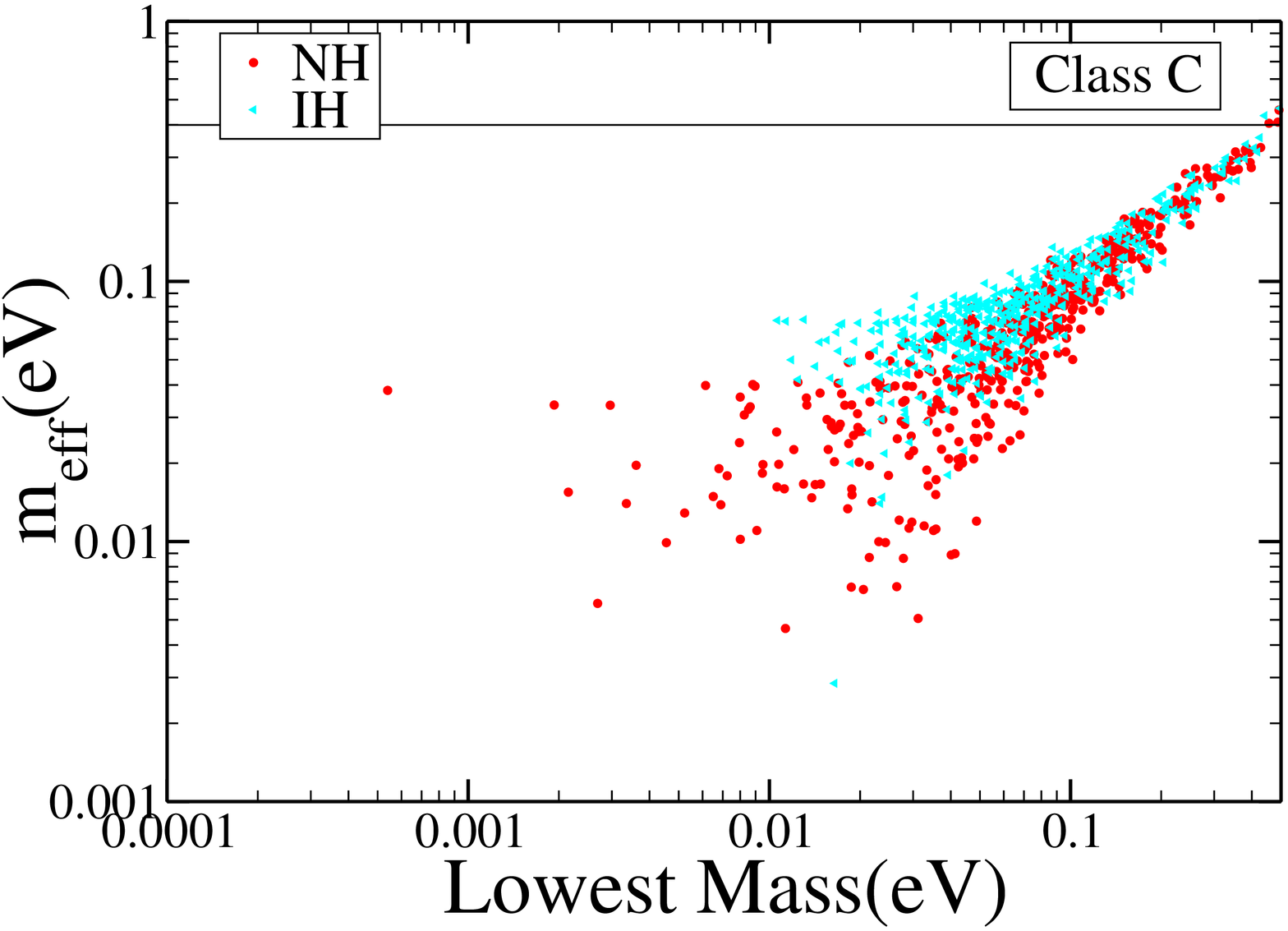} \\
\includegraphics[width=0.25\textwidth,angle=270]{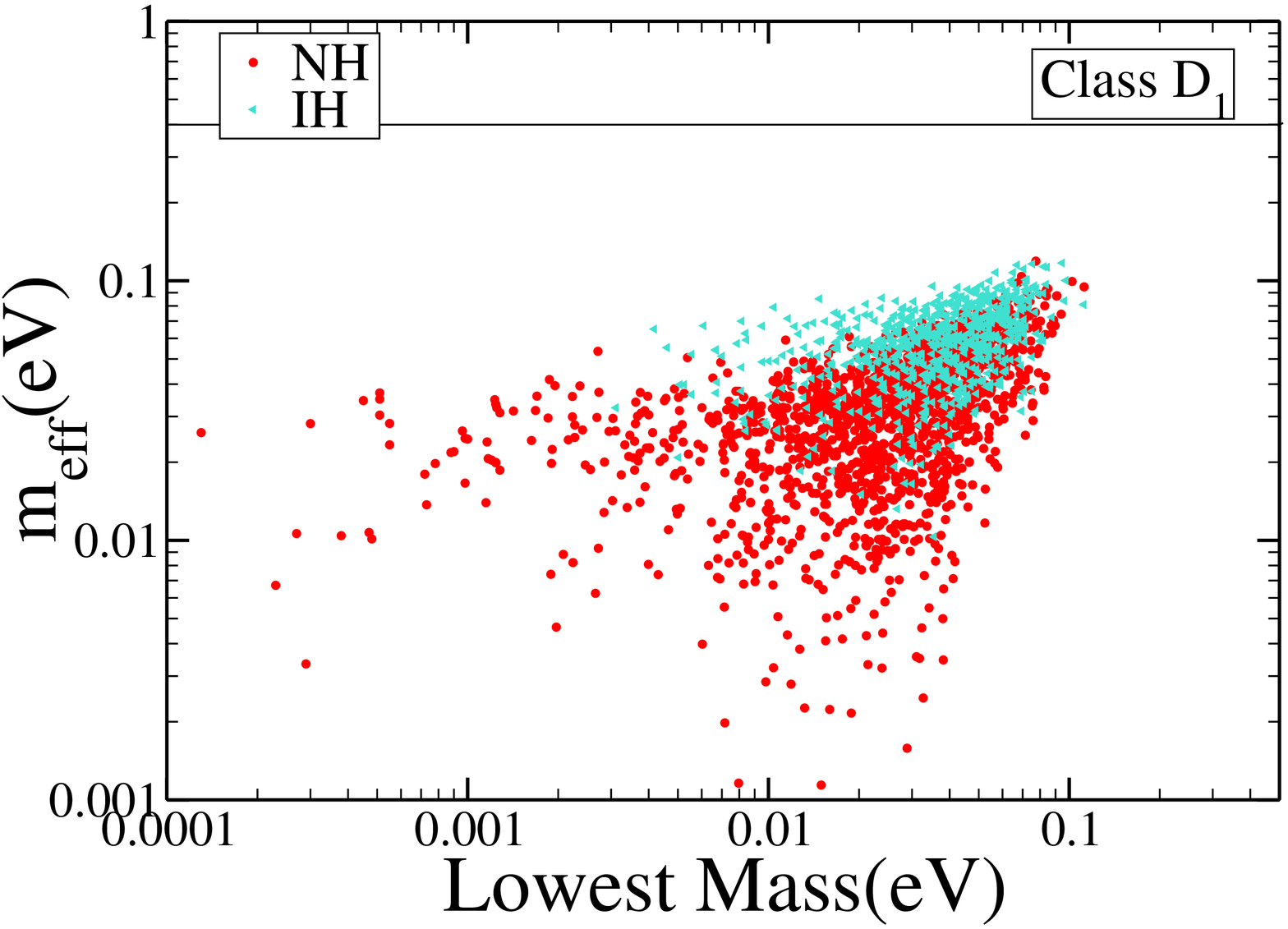}
\includegraphics[width=0.25\textwidth,angle=270]{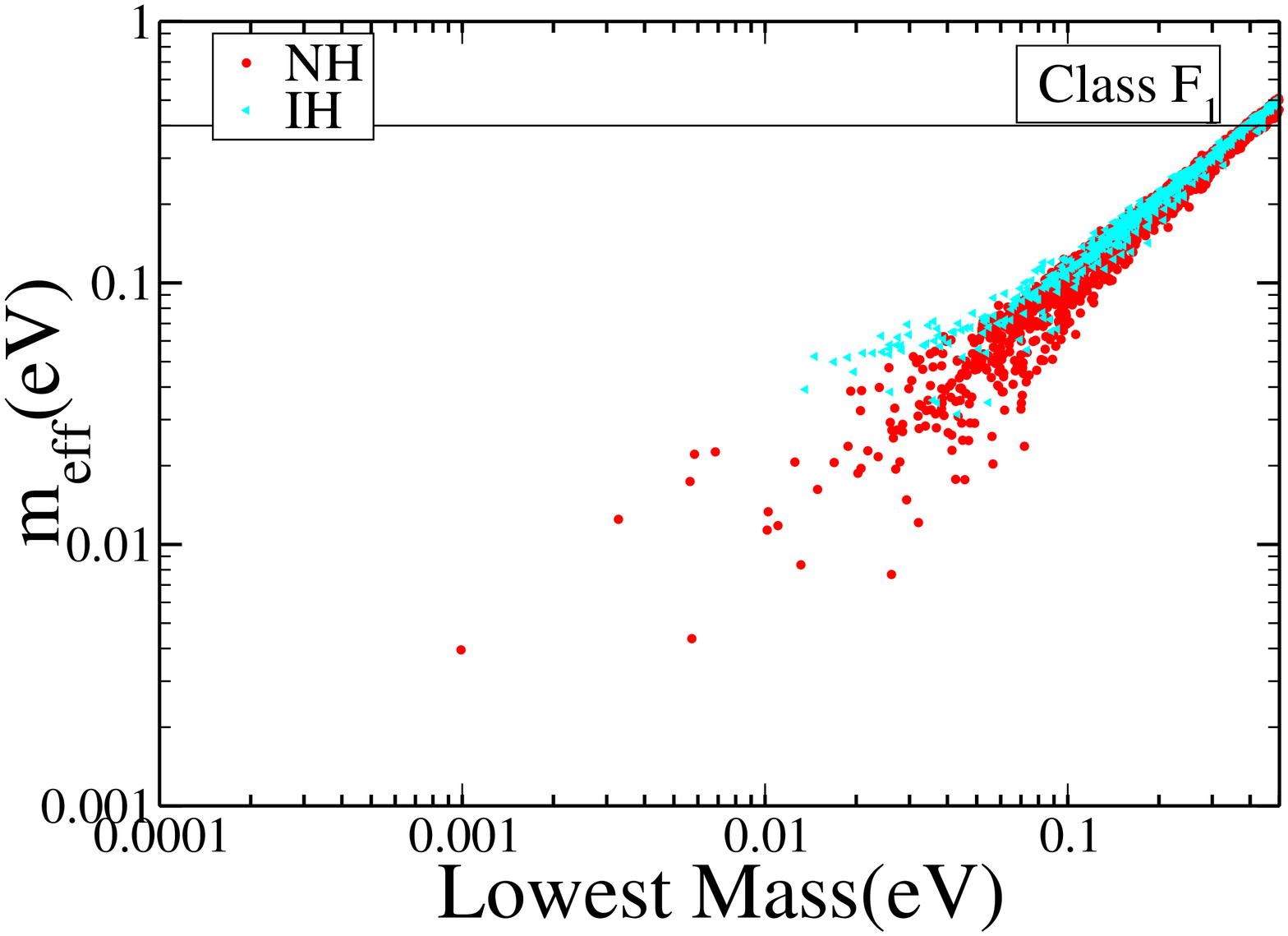}
\includegraphics[width=0.25\textwidth,angle=270]{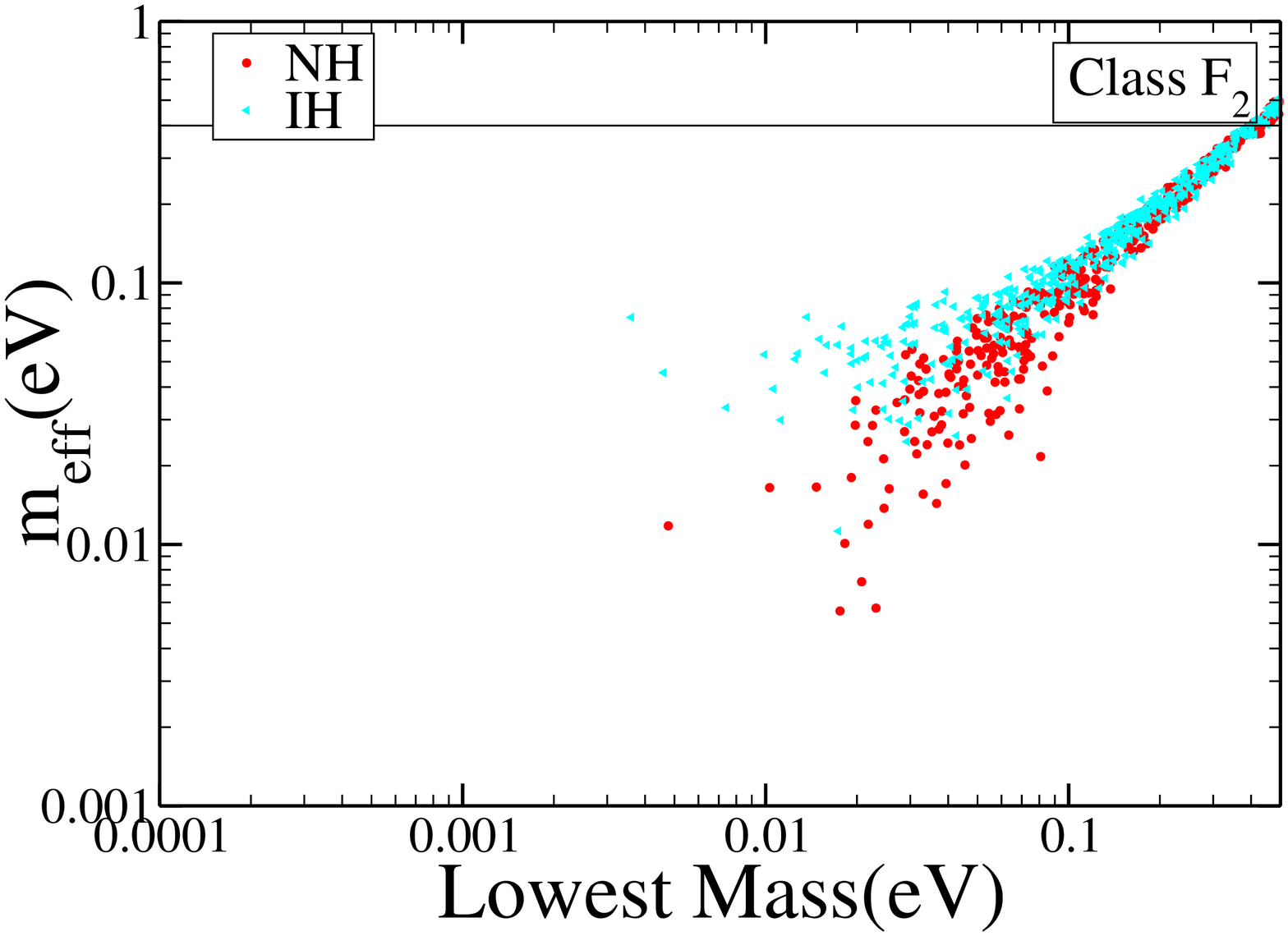}
\begin{center}
Figure 4: The effective mass governing $0\nu\beta\beta$ as a function of 
the lowest mass. The red (dark) points correspond to NH while the cyan (light) 
points correspond to IH.  
\end{center}
\label{fig:effm}
\end{figure}

\item{\underline{\textbf{$B, C$  Classes}}}

In the 3+1 scenario, B and C classes
allow all three mass spectra -- NH, IH and 
QD  assuming the known oscillation parameters to be normally distributed 
(cf Table 3).
In this case since
$\theta_{23}$ in the textures related by 
$\mu-\tau$  
symmetry is not correlated in a simple 
way, the value of this angle not being in the higher octant does not play 
a significant role as in the 3 generation case. 
Among these only $B_1$ allow few points for smaller values of 
$m_1$ for NH. For the IH solution, larger number of points are obtained 
corresponding to the lowest mass $>$ 0.01 eV as is seen from Fig.2. 
In these textures, for higher values of the lowest mass the 
active neutrino contribution  
to the matrix elements are larger and  it is easier to obtain cancellations.
Hence, textures belonging to these classes 
show a preference for QD solutions.
For these textures the  effective mass
governing  $0\nu\beta\beta$ is non-zero.
In the first row of Fig.4 we 
present the effective
mass as a function of the lowest mass for the textures 
$B_1$, $B_3$ and C 
for both NH and IH. These two merge at higher values of the lowest mass
corresponding to the QD solution. The effective mass in these textures is $>0.002$ eV for NH 
and $>$ 0.02 eV for IH. If no signal is
seen in future $0\nu\beta\beta$ experiments then large part of the parameter space belonging to these
textures can be disfavoured.

\item{\underline{\textbf{$D, F$ Classes}}}

These two textures are  disallowed in the 3 generation  case. 
However for the 3+1 scenario they get allowed.
NH is admissible in all the textures belonging to these classes.
The reason for this is the following. \\
In the three active neutrino scenario, the neutrino mass matrix 
in a $\mu-\tau$ block has the elements of the order of $\sqrt{\Delta m^2_{23}}\approx {0.01}eV$ for normal hierarchy. 
Thus, in general these elements are quite large and cannot vanish
\cite{frampton}. However, in the 3+1 case when there is one additional
sterile neutrino, the neutrino mass matrix elements get contribution from the sterile part of the form $m_4 U_{k4} U_{l4}$ where $k=e,\mu,\tau$.
This term is almost of the same order of magnitude and thus can cancel
the active part, resulting into the possibility of vanishing elements
in the $\mu-\tau$ block. Thus, 
the zero textures which were 
disallowed for NH  are now allowed by the inclusion of sterile neutrino (3+1 case). 
In the case of IH, $m_{\mu \tau}$ element for three active neutrinos is always of the order of $\sqrt{\Delta m^2_{23}}\approx {0.01}eV$ and
thus the textures $D_1$, $D_2$, $F_2$, $F_3$  which 
requires $m_{\mu \tau} =0$  were not allowed.
However, for the 3+1 scenario the
extra term coming due to the fourth state
helps in additional cancellations and IH gets allowed 
in these (cf. Table 3). For $F_1$ class, IH for three active neutrino 
is disfavoured because of phase correlations. 
However, with the additional sterile neutrino this can be evaded making it 
allowed. 
In the bottom row of Fig.4 we present the effective mass governing
$0\nu\beta\beta$ for the textures $D_1$, $F_1$ and $F_2$ 
as a function of the lowest mass. The texture
$D_1$ allows lower values of 
$m_1$ for NH while for IH the lowest mass is largely $\gsim 0.01$ eV. 
QD solution is not allowed in D class . 
For F class more points are obtained 
in the QD regime. 
Future experiments on $0\nu\beta\beta$ would be able to probe these regions 
of parameter space.

\end{itemize}

The results presented above are obtained
by varying the  known oscillation parameters given in Table 2 
as distributed normally around their best-fit and with a
width given by the 1$\sigma$ range of the parameters.
There is a finite probability of getting the points in the
3$\sigma$ range of this Gaussian distribution although more
points are selected near the best-fit values.
However, note that  for some of the parameters
the 3$\sigma$ range obtained in this procedure
is different from that presented in Table 2. 
Thus, our results may change if we vary the parameters randomly in their 
3$\sigma$ range as we have seen in the 3 generation case. 
In Fig. 5 we show the allowed values of $y_m$ as a function 
of the lowest mass for the case where all the parameters are varied 
randomly in their 3$\sigma$ range. 
We find that lower values of the smallest mass get disfavoured by this method. 
The main reason for this is that if we use the Gaussian method then the 
allowed 3$\sigma$ range of the mixing angle $s_{14}^2$ is 
from (0.002 - 0.048)  while that of $s_{24}^2$ is from (0.001 - 0.06). 
Thus, smaller values of $s_{14}^2$ and $s_{24}^2$  are possible which
helps in achieving cancellation conditions for smaller values of
$m_1$ or $m_3$.
But if the parameters are varied randomly in the 3$\sigma$ range 
presented in Table 2
then such smaller values 
of the angles  are not allowed and consequently 
no allowed points are obtained for smaller values of masses. 
In particular, we obtain 
$m_1$(NH) or $m_3$(IH) $>$ 0.01 eV  in all the textures 
However, 
main conclusions presented in Table \ref{2results}
regarding the nature of the allowed mass spectrum for the 3+1 scenario remain
unchanged though the allowed parameter space gets reduced.
Specially for the 
A, E and C  classes very few points  get allowed.
Fully hierarchical neutrinos ($m_1 < m_2 < m_3$) are not 
possible in any of the textures. Textures belonging to 
the B and F classes give more points in the 
QD regime. D class allows partial NH or IH.

\begin{figure}
\begin{center}
\includegraphics[width=0.25\textwidth,angle=270]{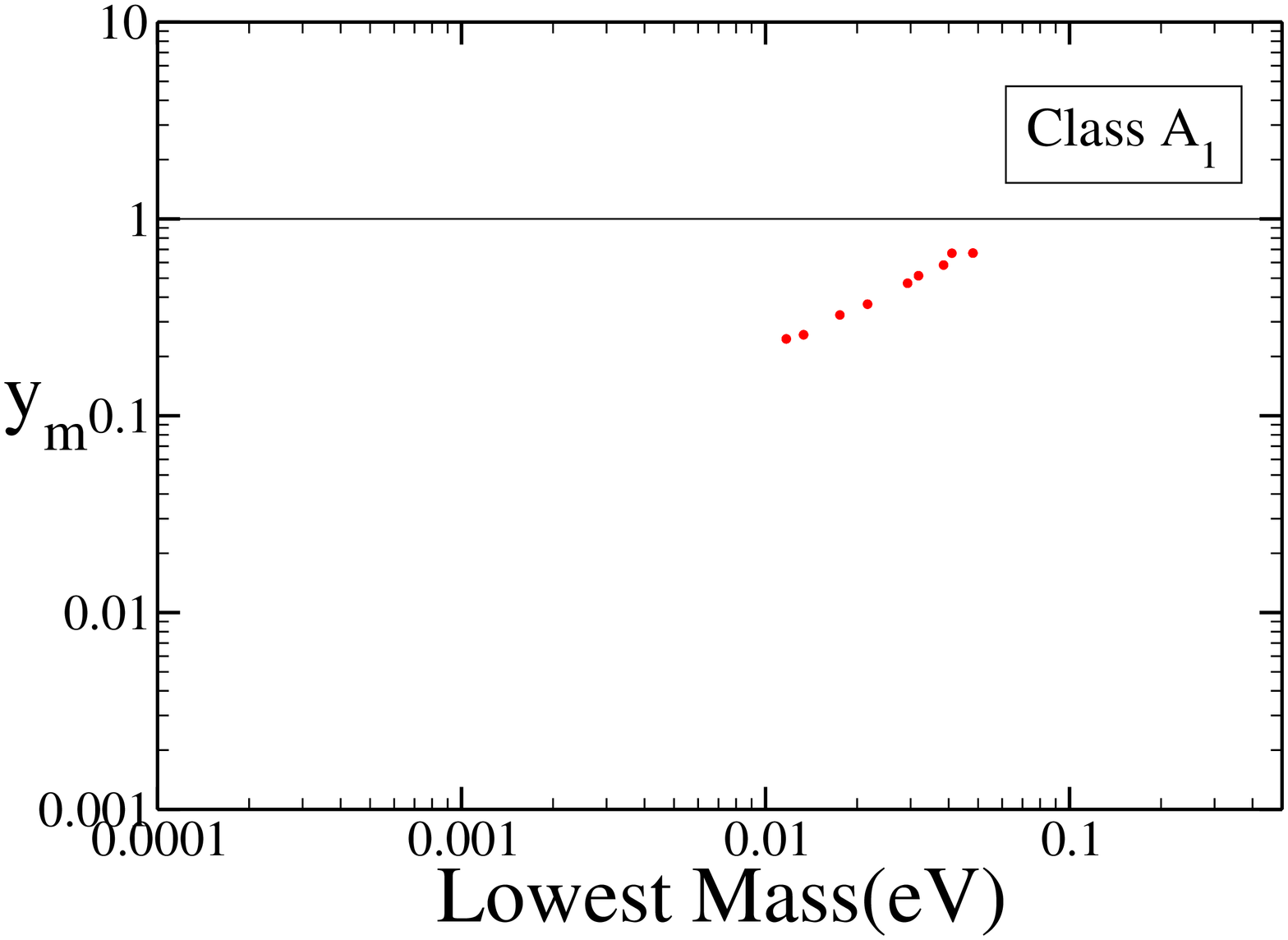}
\includegraphics[width=0.25\textwidth,angle=270]{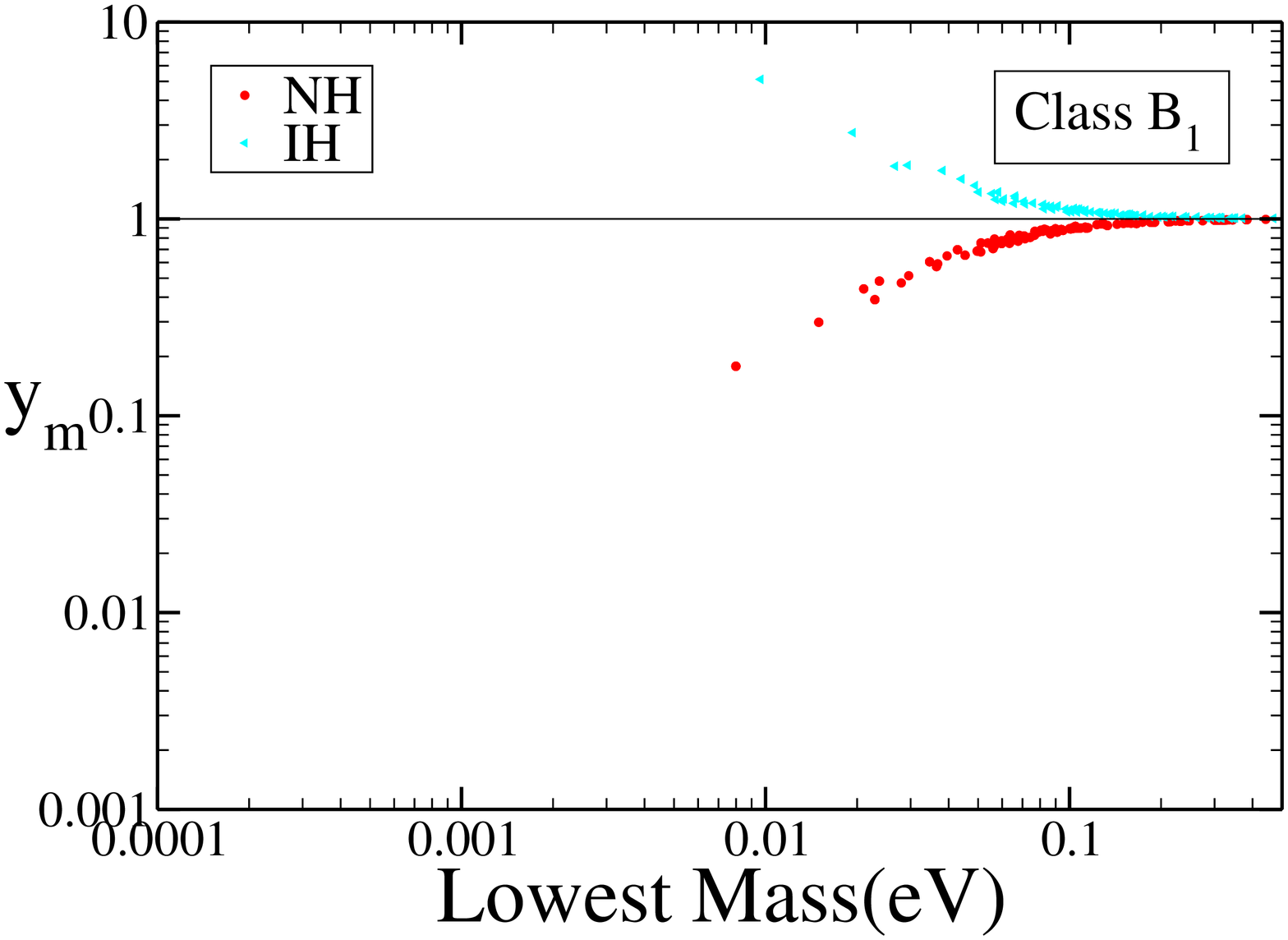}
\includegraphics[width=0.25\textwidth,angle=270]{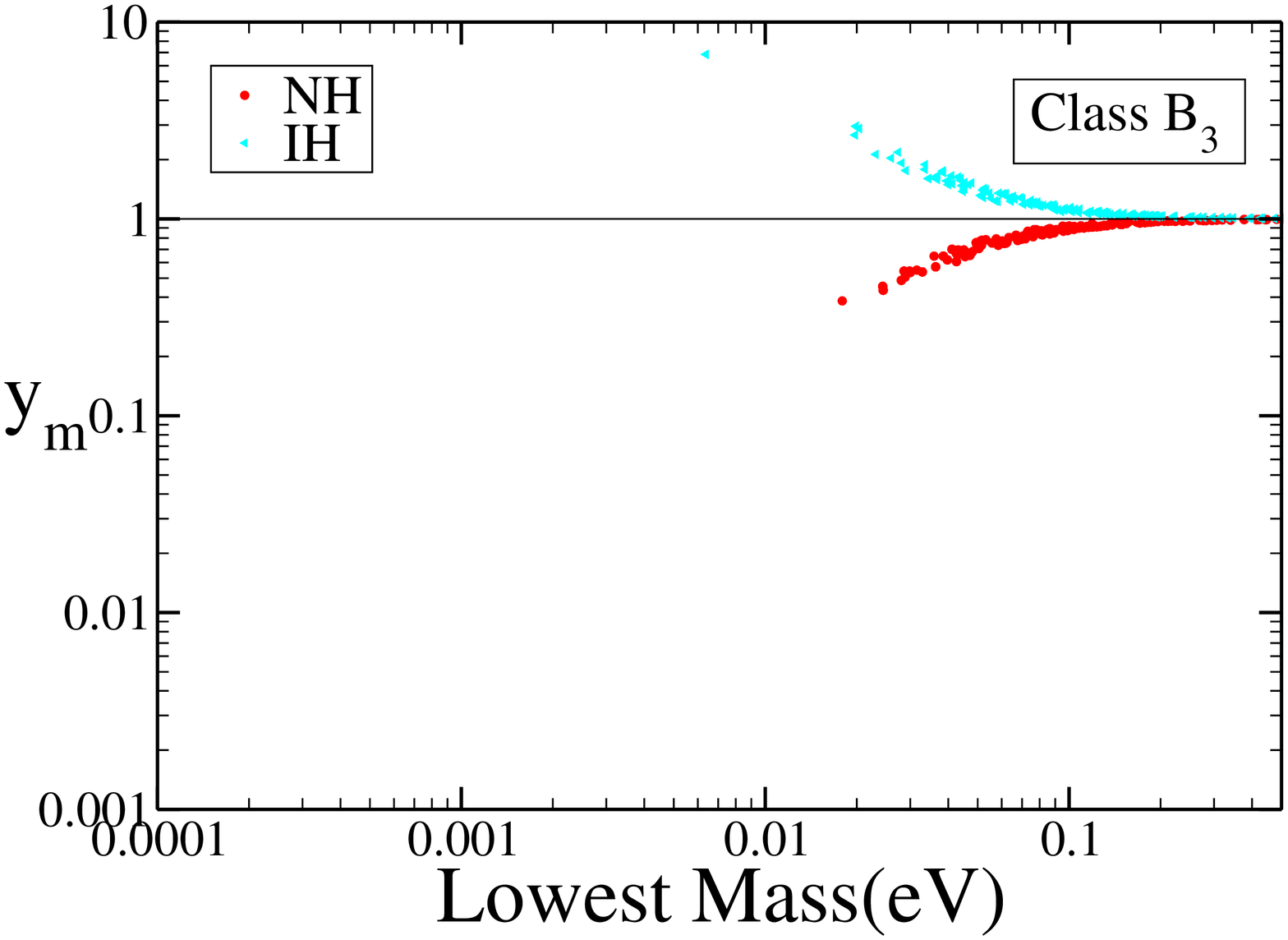} \\
\includegraphics[width=0.25\textwidth,angle=270]{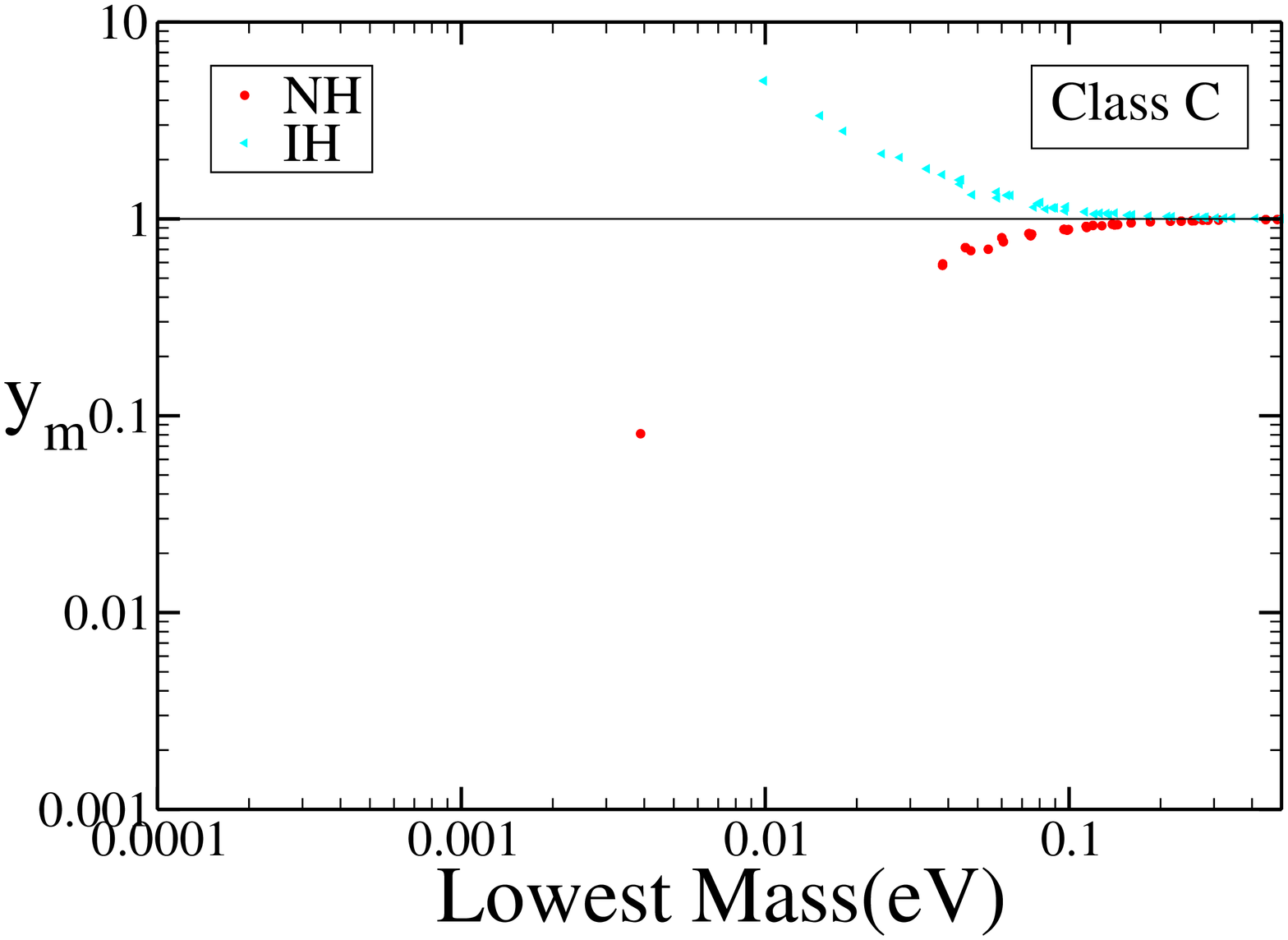}
\includegraphics[width=0.25\textwidth,angle=270]{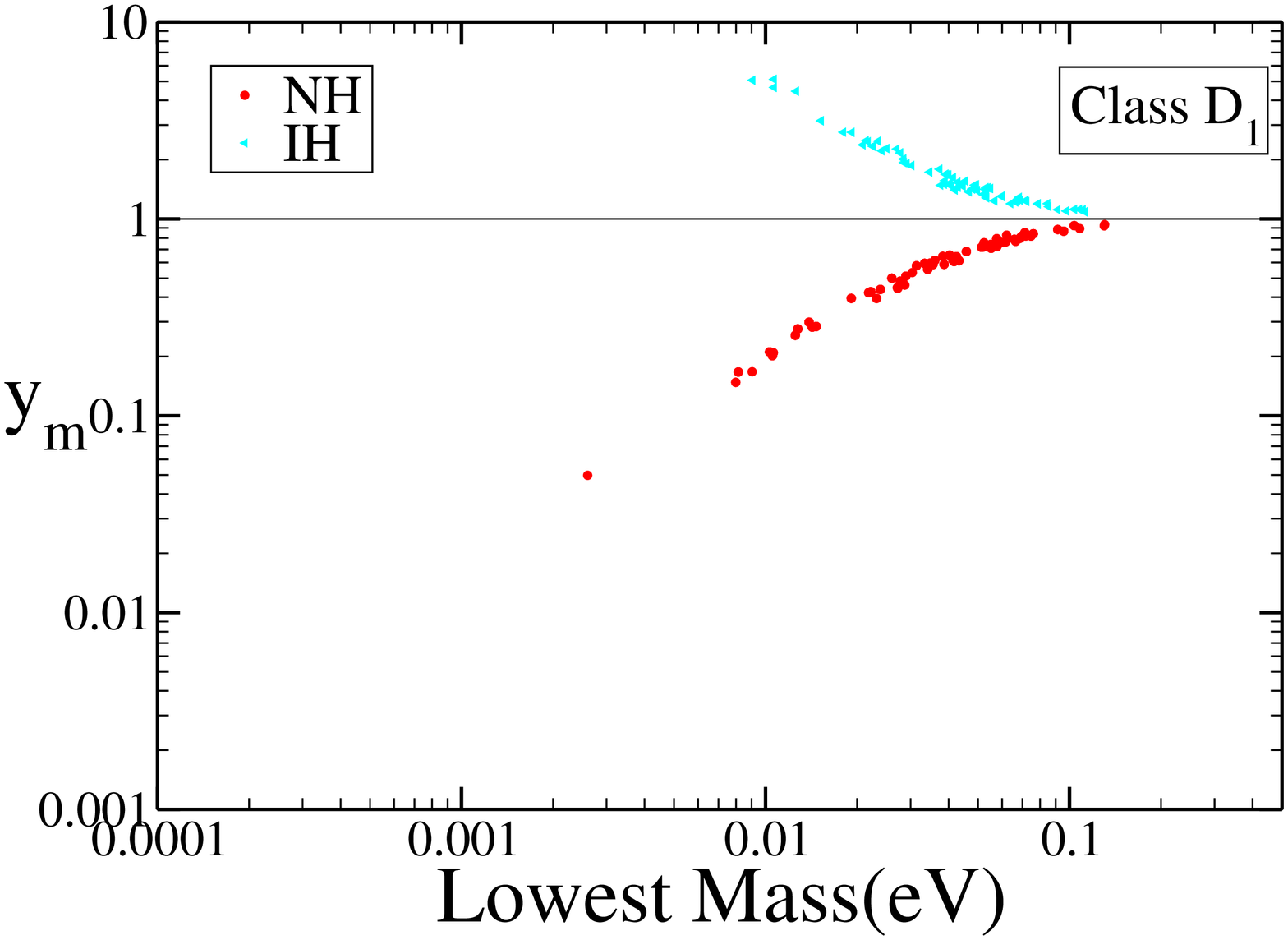}
\includegraphics[width=0.25\textwidth,angle=270]{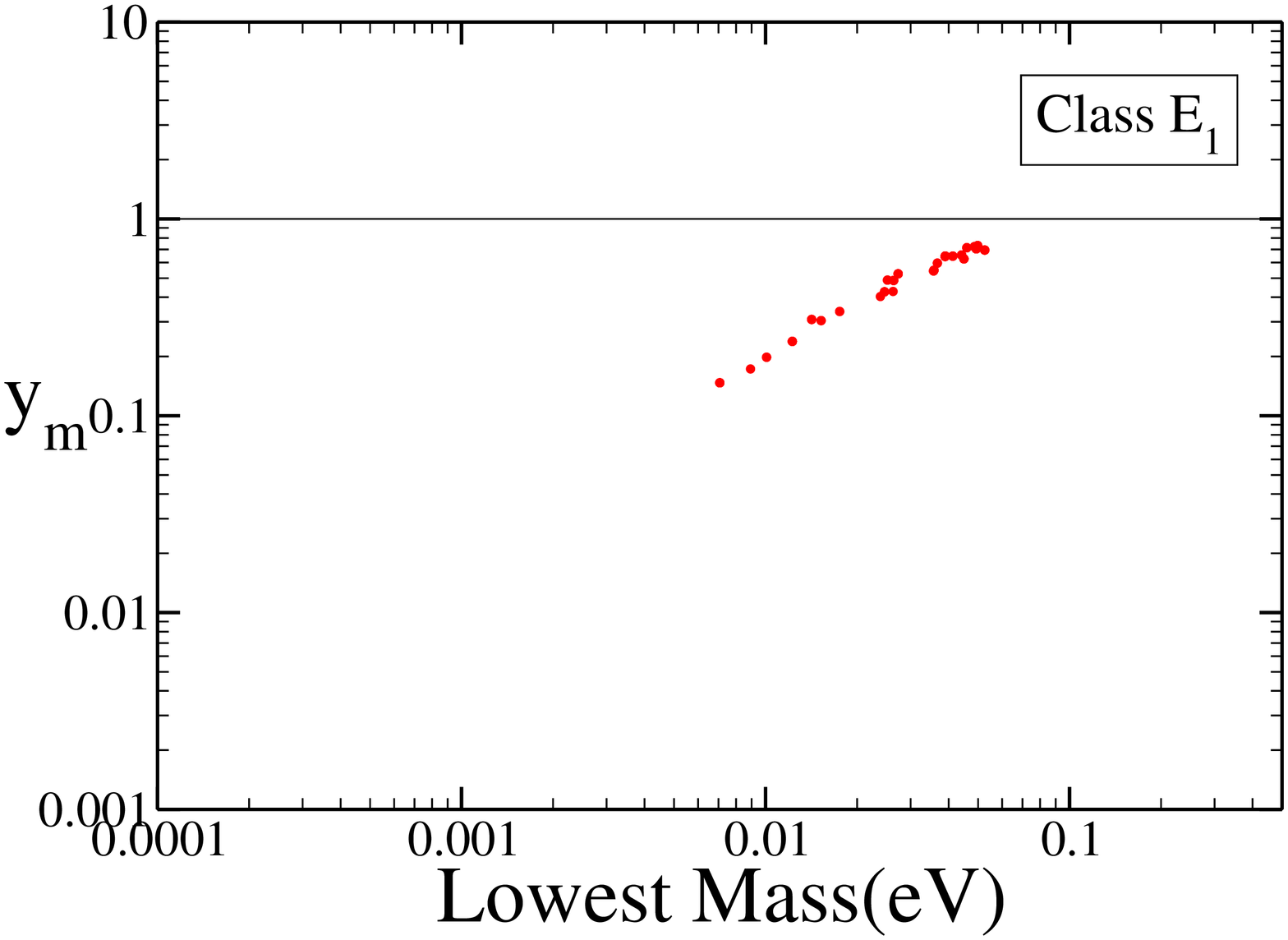} \\
\includegraphics[width=0.25\textwidth,angle=270]{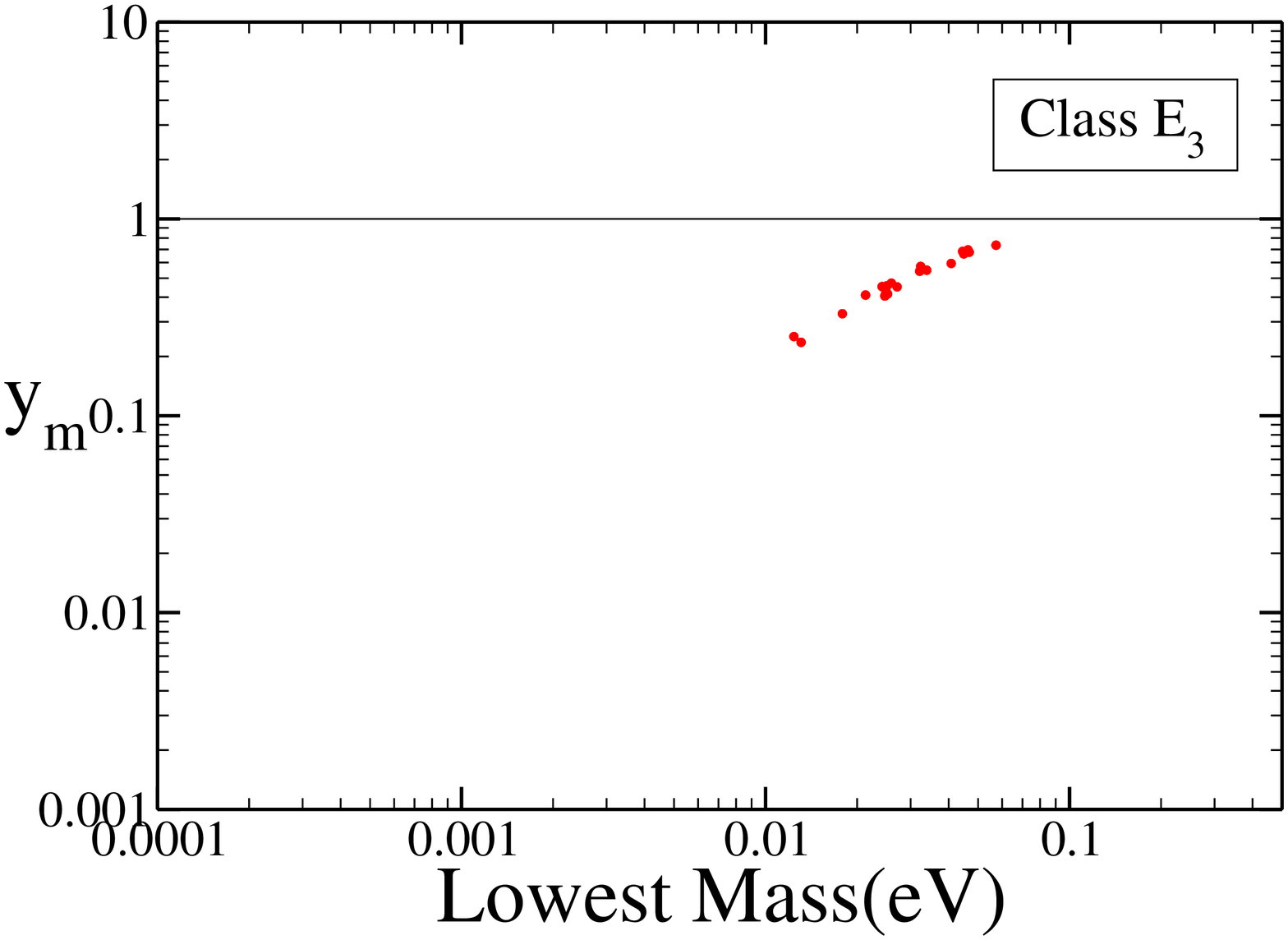}
\includegraphics[width=0.25\textwidth,angle=270]{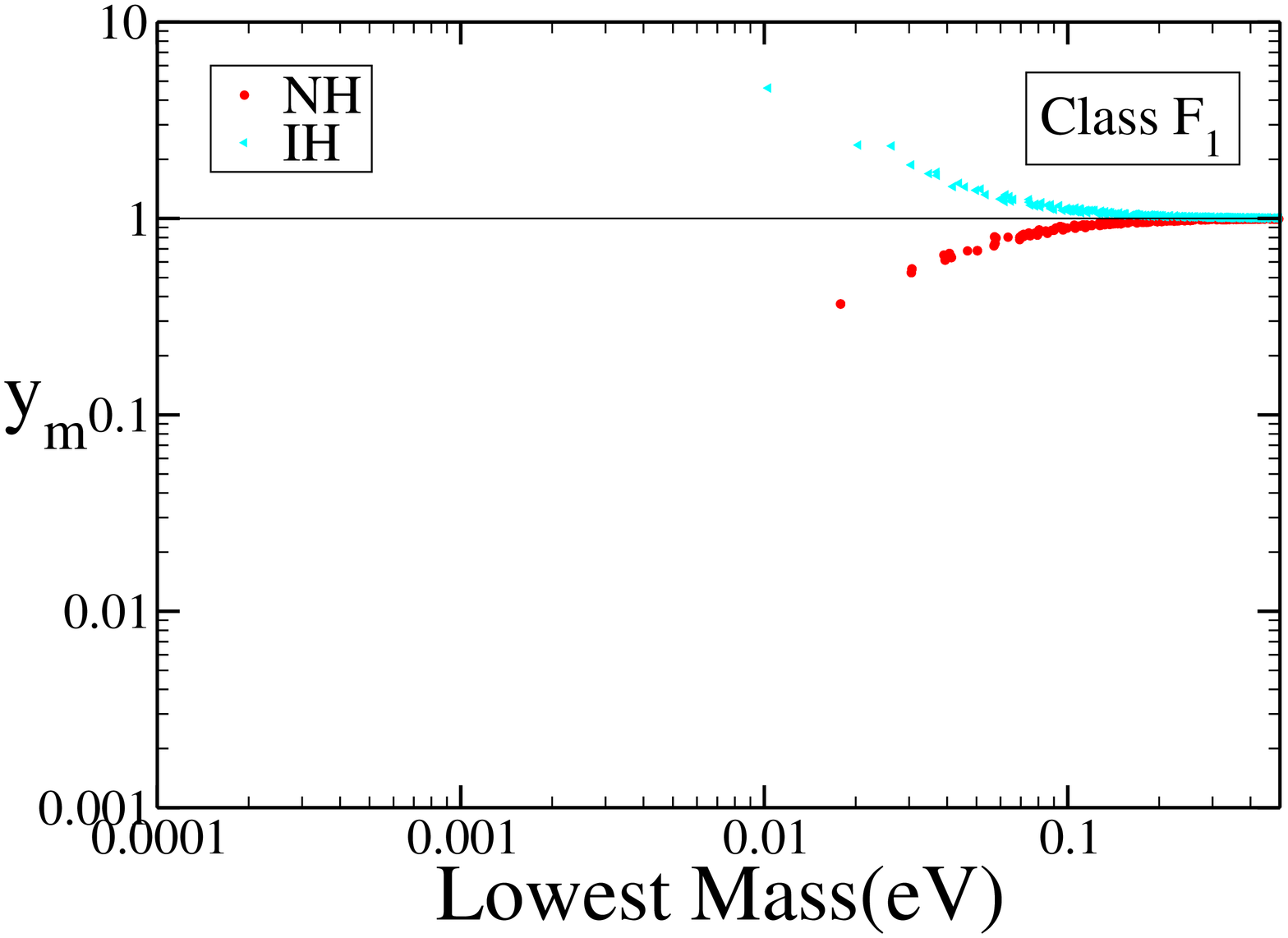}
\includegraphics[width=0.25\textwidth,angle=270]{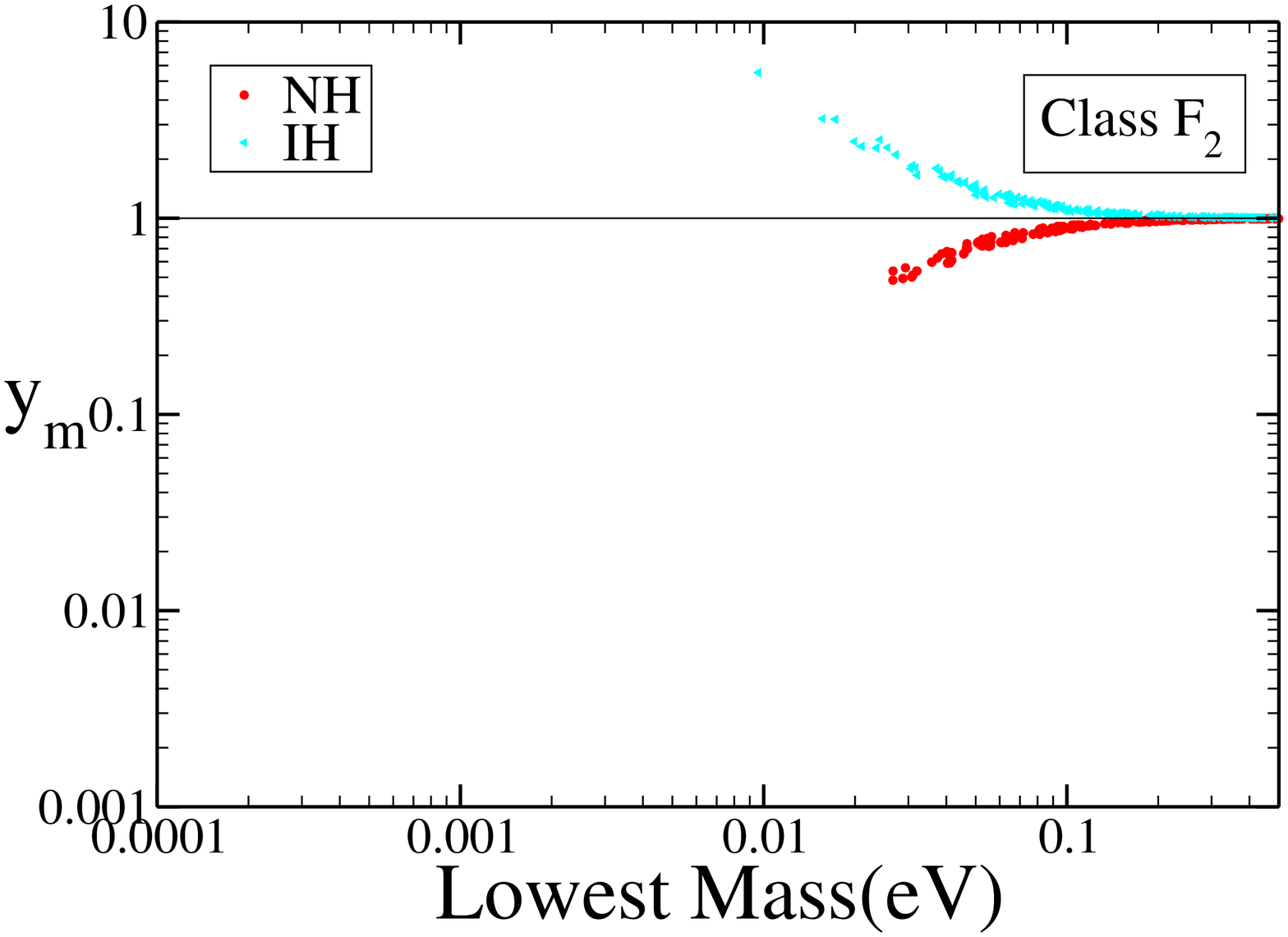}
\end{center}
\begin{center}
Figure 5: The values of $y_m$ as a function of
the lowest mass when parameters are varied randomly.
\end{center}
\label{yrandom} 
\end{figure}

\section{Conclusions}

Recent experimental observations make a case for enlarging the
scope of three flavour oscillations to include
one or more sterile neutrinos with mass around $\sim$ 1 eV.
Although induction of
more than one sterile neutrino may provide a better fit to the
oscillation data the cosmological observations may  be
more consistent with
the three active and one sterile picture with the sterile neutrino
being heavier. With the  addition of one sterile neutrino the
parameter space describing neutrino masses and mixing at low energy
increases to include
four independent masses, six mixing angles and six phases.
The low  energy mass matrix in the flavour basis now consists of 10
independent elements as opposed to six elements for the three generation
case. It is well known that for the three generation case the neutrino
mass matrix in flavour basis can have at the most two zeros.
In this work, we have considered the two zero mass matrices in the framework
of three active and one sterile neutrino.
We find many distinctive features in this case as compared to the three
neutrino scenario. For the 3+1 case there can be 45 possible two
zero textures as opposed to 15 for the 3 generation case.
Among these 45 possible two-zero textures only 15 survive the
constraints from global oscillation data. Interestingly these
15 cases are the 15 two-zero textures that are possible
for three active neutrino mass matrices. While for the
three active neutrino case only 7 of these were allowed
addition of one sterile neutrino make all 15 cases allowed
as the sterile contribution can be instrumental for additional
cancellations leading to zeros.
All the allowed textures admit NH.
The classes B, C, F also allow
IH and QD solutions in addition.
The results are summarized in Table \ref{2results}. 

If we vary the mass and mixing parameters normally peaked at the best-fit value and 1$\sigma$ error as the width then we find solutions for smaller 
values of $m_1$(NH) and $m_3$(IH). 
In this case 
for the textures with $m_{ee}=0$ i.e A class and E class
we obtain correlations between the mixing angles $\sin^2\theta_{14}$,
$\sin^2\theta_{24}$, $\sin^2\theta_{34}$
and the lowest mass scale $m_1$. For these textures the
effective mass responsible for neutrinoless double beta decay is zero.
For the other allowed textures
we present
the effective mass measured in
neutrinoless double beta decay as a function of the smallest
mass scale.
If however, the known oscillation parameters 
are varied randomly in their allowed
3$\sigma$ range  then although the main conclusions 
deduced above regarding the allowed mass spectra in various
textures remain the same  the allowed parameter space 
reduces in size. In particular,  we obtain a
bound on the smallest mass as $m_{\rm smallest} > 0.01$ eV 
and completely hierarchical neutrinos are no longer allowed.  

In this work we have concentrated on the two zero textures. 
However it is possible that the neutrino mass matrix in the 3+1 picture
may allow the presence of more than two zeros \cite{workinprogress}.
These results can be useful in probing underlying flavour symmetries and
also for  obtaining textures of Yukawa matrices
in presence of light sterile neutrinos should their existence be confirmed in
future experiments. 

\section{Acknowledgements}

The authors thank Werner Rodejohann for his involvement in the
initial stages of the work and many useful discussions.
They would also like to thank Anjan Joshipura for helpful
discussions.

\end{document}